# Spiral Spin Liquid Noise

Hiroto Takahashi[1]†, Chun-Chih Hsu[1]†, Fabian Jerzembeck[1,2], Jack Murphy[3], Jonathan Ward[3], Jack D. Enright[1,3], Jan Knapp[1], Pascal Puphal[4], Masahiko Isobe[4], Yosuke Matsumoto[4], Hidenori Takagi[4], J. C. Séamus Davis[1,2,3,5]*, Stephen J. Blundell[1]*

1. Clarendon Laboratory, University of Oxford; Oxford, OX1 3PU, UK.
2. Max Planck Institute for Chemical Physics of Solids; Dresden, D-01187, DE.
3. Department of Physics, University College Cork; Cork, T12 R5C, IE.
4. Max Planck Institute for Solid State Research; Stuttgart, D-70569, DE.
5. Department of Physics, Cornell University; Ithaca, NY 14853, USA.

† These authors contributed equally to this work.

* Corresponding authors. Email: jcseamusdavis@gmail.com & Stephen.Blundell@physics.ox.ac.uk

ABSTRACT An emerging concept for identification of different types of spin liquids[1,2,3] is through the use of spontaneous spin noise[4,5,6,7,8,9]. Here we develop spin noise spectroscopy for spin liquid studies by considering $Ca_{10}Cr_7O_{28}$, a material hypothesized to be either a quantum or a spiral spin liquid. By enhancing techniques introduced for magnetic monopole noise studies[10,11,12] we measure the time and temperature dependence of spontaneous flux $\Phi(t,T)$ and thus magnetization $M(t,T)$ of $Ca_{10}Cr_7O_{28}$ samples. The resulting power spectral density of magnetization noise $S_M(\omega,T)$ reveals intense spin fluctuations with $S_M(\omega,T) \propto \omega^{-\alpha(T)}$ and $0.84 < \alpha(T) < 1.04$. Both the variance $\sigma_M^2(T)$ and the correlation function $C_M(t,T)$ of this spin noise undergo crossovers at a temperature $T^* \approx 450\,\text{mK}$. While predictions for quantum spin liquids are inconsistent with this phenomenology, those from Monte-Carlo simulations of a 2D spiral spin liquid state in $Ca_{10}Cr_7O_{28}$ yield overall quantitative correspondence with the measured frequency and temperature dependences of $S_M(\omega,T)$, $C_M(t,T)$ and $\sigma_M^2(T)$, thus indicating that $Ca_{10}Cr_7O_{28}$ is a spiral spin liquid.

In theory, spin liquids can occur in either classical[1] or quantum[2,3] incarnations. The former exhibits massive ground-state degeneracy of its spin configurations, while the latter exhibits quantum entanglement of its localized spins along with fractionalized spin excitations. Candidate materials for either of these states are most often designated based on the absence of long-range magnetic order when the energy scale of magnetic interactions greatly exceeds that of temperature[1,2,3]. But no spin liquid can be identified conclusively in this way, meaning that innovative techniques are urgently required to specify each material's spin liquid state. A promising new concept is to measure the unique spectrum of spontaneous spin noise generated by quantum and thermal fluctuations, thereby 'fingerprinting' each spin liquid state[4,5,6,7,8,9]. For example, in fermionic atomic vapours, the variance of magnetization-noise $\sigma_M^2 \equiv \langle M^2(t) \rangle$ distinguishes the Bardeen-Cooper-Schrieffer superfluid state from the Bose-Einstein-Condensate state and from the antiferromagnetic state[4]. Or, for the case of random exchange coupling Heisenberg spin-1/2 phases, theory predicts magnetization noise exhibiting $S_M(\omega) \propto \omega^{-\alpha}$ with $0.5 < \alpha < 1$ due to finite temperature many-body-localization[6]. Finally, for a canonical $U(1)$ gapless quantum spin liquid state with a spinon Fermi surface, theory predicts white magnetization noise for which $S_M(\omega)$ is a constant[8,9].

The utility of this approach has recently been demonstrated[10,11,12,13,14] for the case of emergent magnetic monopoles[15,16,17] in spin-ice e.g. $Dy_2Ti_2O_7$ and $Ho_2Ti_2O_7$; and in artificial spin ices. There, thermally activated spin flips generate emergent magnetic monopole charges $\pm m$ . Generation-recombination theory for these monopoles then predicts magnetization noise $S_M(\omega, T) = 4\sigma_M^2(T)\tau(T)/(1 + (\omega\tau(T))^2)$ where $\omega$ and $T$ are angular frequency and temperature, $\sigma_M^2(T)$ is the magnetization variance and $\tau(T)$ is relaxation time[18]. Congruently, Monte Carlo (MC) simulations from the realistic spin-ice Hamiltonian[19] predict a closely related magnetization noise spectrum $S_M(\omega, T) \propto \tau(T)/(1 + (\omega\tau(T))^b)$ where $b(T) < 2$ because of correlations in the monopole motion[10, 20]. Most recently, discovery of the dynamical fractal nature of monopole trajectories[12] yielded the prediction that $S_M(\omega, T) \propto \tau(T)/(1 + (\omega\tau(T))^b)$, with $b(T) = 1.5$ because there are two different possible microscopic spin-flip rates. By now, virtually all key predictions of spin noise theories[10,12,18] specific to the monopole dynamics in spin-ice pyrochlores have been borne out directly in SQUID-based spin noise spectroscopy experiments[10,11,13,14] that measure $S_M(\omega,T)$ and correlation function $C_M(t,T)$ of $Dy_2Ti_2O_7$. Evidently, these achievements provide motivation to deploy this technique more generally for spin liquid research.

To do so we study $Ca_{10}Cr_7O_{28}$[21,22,23,24,25], a quasi-2D material consisting of weakly coupled bilayers (Fig. 1A). Each is a buckled kagome lattice in which the triangular plaquettes have alternating sizes[22]. The magnetic $Cr^{5+}$ ions have a six-site unit cell, with each $Cr^{5+}$ located within a distorted $CrO_4$ tetrahedron and having a singly occupied $S = 1/2$ state (Fig. 1A). The isotropic magnetic susceptibility of $Ca_{10}Cr_7O_{28}$ exhibits a Curie-Weiss temperature $T_{\mathrm{CW}} = +2.35$ K indicative of ferromagnetic interactions on the $J \lesssim 1$ meV energy scale[23]. The zero-field magnetic specific heat capacity $C(T)$ has a broad maximum at $T \approx 3$ K with a sharper peak followed by precipitous drop below $T^* \approx 450$ mK [21]. For $T < T^*$ it exhibits an approximately linear temperature dependence $C(T) \cong \eta T$ [25]. Zero-field muon spin rotation measurements exclude magnetic order down to $T \approx 20$ mK and instead evidence spin fluctuations which slow down on cooling and become persistent below $T^*$ [21]. Inelastic neutron-scattering detects a spin excitation spectrum $\Sigma(\boldsymbol{q}, E)$ lacking well-defined spin-wave modes[25] but with a ring-like closed contour of scattering at intermediate energies[21,23,25].

Two distinct spin liquid scenarios have been envisioned to explain the $T < T^*$ state. The first hypothesis is that $Ca_{10}Cr_7O_{28}$ is a quantum spin liquid (QSL). No magnetic order whatsoever is detected at temperatures down to $T \approx 20$ mK, thermodynamically, by elastic/inelastic neutron scattering, or by muon spin precession. The relaxation rates of muon polarization $P(t)$ demonstrate that the spins remain entirely dynamic to the same low temperature[21]. A continuum of dispersionless spin excitations exists along with a diffuse ring of intense scattering at $\hbar\omega \sim 0.3$ meV, all confined to the kagome plane. Further, based on the linear temperature dependence of magnetic specific heat, a QSL with $Z_2$ symmetry and a spinon Fermi surface has been inferred[25]. Finally, pseudo-fermion functional renormalization group theory[21] or tensor network theory [26] both indicate a quantum magnetic ground state. The consequent principal hypothesis has been that $Ca_{10}Cr_7O_{28}$ is a QSL[21,22,23,24,25,26]. On the other hand, it has been conjectured that $Ca_{10}Cr_7O_{28}$ is a spiral spin liquid (SSL). In this state[27-40] a spin spiral occurs at wavevector $\boldsymbol{Q}$, during each modulation of which the spin direction undergoes a spiral evolution (Fig. 1B). In a SSL the vectorial direction $\boldsymbol{Q}$ is not fixed but occupies a continuous closed contour[27,35,37] in reciprocal space (Fig. 1B). Such systems with a sub-extensive degeneracy avoid long-range ordering[41] and are distinct from any static magnetically ordered state with well-defined spin wave modes. The SSL state has been mainly studied in a classical context and is typically conceived to be in a classical spiral spin liquid state. Accordingly, we consider the classical situation throughout this paper, although we note that the interplay of quantum effects and the degenerate spiral manifold remains an open question as a non-ordered 'quantum spiral spin liquid' is a theoretical possibility[42,43]. The $Ca_{10}Cr_7O_{28}$ model of spin-1/2 on a distorted bilayer kagome can be mapped to interacting spin-3/2 on a monolayer honeycomb lattice (Fig. 1A), when 3 spins on alternative triangular plaquettes form a $S$ = 3/2 state by ferromagnetic interactions[25,33,34] so that the expected spin behaviour is closer to the classical limit. This model is frustrated and its MC simulation predict a SSL with $q$ -space ring-like correlations[33,34] consistent with the experiment[21,23,25]. Generic 2D XY models for the SSL state further predict the existence of a unique topological defect referred to as a momentum vortex[37] which could dominate the low-energy physics. This defect occurs at a point around which the SSL continuously occupies all possible $\boldsymbol{Q}$-vector states on its manifold (Fig. 1C). MC simulations predict that, as a result of the nonlocality of these topologically constrained momentum vortices, dynamics slows as temperature is decreased and eventually reaches a metastable configuration. Consequently, the evolution is from a trivial paramagnet into a 'pancake' liquid state and eventually into a SSL before freezing into a vortex lattice[37]. Thus, on the basis of a non-ordered dynamic spin state having diffuse spin correlations seen in a closed contour of more intense neutron scattering, $Ca_{10}Cr_7O_{28}$ is hypothesized to be a SSL[33,34,37]. Based on the extant phenomenology of $Ca_{10}Cr_7O_{28}$, however, it has not been possible to distinguish conclusively between these QSL and SSL scenarios.

To address this challenge using techniques introduced here, the spin noise spectra of a QSL and a SSL state are required. For $Z_2$ and $U(1)$ QSL, Ref. 8 predicts frequency-independent noise power spectral density $S(\omega) \propto \omega^0$ in the high-temperature limit $\omega \ll T$ where our studies are carried out (Methods). For SSL on the other hand rapid theoretical advances[37] have occurred recently using MC simulations based on a generic XY model for a 2D SSL with Hamiltonian

$$H = J_1 \sum_{<ij>1} \boldsymbol{\vartheta}_i \cdot \boldsymbol{\vartheta}_j + J_2 \sum_{<ij>2} \boldsymbol{\vartheta}_i \cdot \boldsymbol{\vartheta}_j + J_3 \sum_{<ij>3} \boldsymbol{\vartheta}_i \cdot \boldsymbol{\vartheta}_j \quad (1)$$

Here $\boldsymbol{\vartheta}_i$ represents XY spins constrained to lattice sites in a plane, and $J_1; J_2; J_3$ are the first, second, and third nearest neighbor spin couplings. For parameters $J_1 = -1; J_2 > 1/4; J_3 = J_2/2$ this system is a SSL exhibiting a ground-state set of spiral density waves with momenta $\boldsymbol{Q}$ satisfying $2\cos^2 Q_x + 2\cos^2 Q_y + 4\cos Q_x \cos Q_y = 1/2J_2^2$ : this is a continuous closed contour in $q$-space within the first reciprocal unit cell of the lattice (Fig. 1B). The full details of our MC simulations are presented in Methods using equation (1) with $J_1 = -1.0; J_2 = 0.28; J_3 = 0.14$ for an array of $N = L \times L$ spins on a square lattice with periodic boundary conditions. Such semiclassical simulation results using $S = 3/2$ for $Ca_{10}Cr_7O_{28}$ are well supported by successful previous studies[33,34] (Methods). Equilibration of $\boldsymbol{\vartheta}(\boldsymbol{r}, T)$, the in-plane spin vector at each site $\boldsymbol{r}$, to temperature $T$ uses an initial set of $\boldsymbol{\vartheta}(\boldsymbol{r})$ randomly selected with uniform probability. Examples of $\boldsymbol{\vartheta}(\boldsymbol{r}, T)$ evolution in the equilibration process can be visualized in Supplementary Movie 1 and Movie 2. At each temperature this yields a representative spin configuration, four typical examples of which are shown for different $T$ in Fig. 2A. The structure factor $\Sigma(\boldsymbol{q})$ of each spin configuration is shown in Fig. S11. This is in excellent agreement with the evolution from a paramagnet into a pancake liquid state, thence to a SSL and finally into a vortex lattice as reported by Ref. 37.

Such thermalized configurations (Fig. 2A) subsequently initiate the simulation of $\boldsymbol{\vartheta}(\boldsymbol{r}, t, T)$ via the Metropolis Monte Carlo algorithm, which approximates the evolution of spins interacting with a thermal bath i.e. thermal fluctuations of spins. Here, the system is evolved sequentially through $10^7$ MC time-steps at each $T$. During MC simulations at $T \sim 0.15|J_1|$ we find the elementary local spin relaxation process occurs at a timescale ~10 MC step (Methods and Fig. S3). High-frequency AC susceptibility experiments yield a semicircular Cole-Cole plot indicating a microscopic spin relaxation time corresponding to $\sim 10\ \mu$s [21]. Hence setting each MC step to $\tau = 1\ \mu$s simulates a microscopic relaxation time consistent with empirical observations; we then use a total MC simulation run time $\Gamma = 10$

s. From these data, we predict the spin noise fingerprint of the SSL by calculating the $x, y$-components of the average spin $\bar{\vartheta}_{x,y}(t,T) = \frac{1}{N}\sum_{\boldsymbol{r}} \vartheta_{x,y}(\boldsymbol{r},t,T)$ versus time, with typical examples shown in Figs. 2B and S2A. Most importantly, the power spectral density of SSL spin noise is calculated from $S_{\bar{\vartheta}_{x,y}}(\omega_j,T) = \frac{1}{\pi\Gamma}\left|\Delta t \sum_{k=0}^{K-1} e^{-i\omega_j t_k}\bar{\vartheta}_{x,y}(t_k,T)\right|^2$, where $\Delta t$ is time interval and $K = \Gamma/\Delta t$, with typical results shown in Figs. 2C and S2B. With falling temperatures below $T = 0.15|J_1|$, it shows strong low-frequency noise down to at least $\omega/2\pi = 1$ Hz that diminishes continuously in power. The correlation function of this SSL noise is $C_{\bar{\vartheta}_{x,y}}(t_k,T) = \frac{1}{l_{\text{ave}}}\sum_{l=0}^{l_{\text{ave}}-1} \bar{\vartheta}_{x,y}(t_l,T)\,\bar{\vartheta}_{x,y}(t_l+t_k,T)$ where $t_k = k\Delta t$ and $l_{\text{ave}} = 9\times 10^5$; its typical temperature dependence is shown in Figs. 2D and S2C. Well above $T = 0.15|J_1|$, the correlations drop rapidly over time. As temperature falls, the decay of correlation slows down and the functional form becomes nearly $C_{\bar{\vartheta}_{x,y}}(t,T) \propto -\ln t$ (Fig. S10A). To predict the temperature dependence of SSL noise power-law $\alpha(T)$, the $S_{\bar{\vartheta}_{x,y}}(\omega,T)$ is fitted by a function $A(T)\omega^{-\alpha(T)}$ in the frequency range 1 Hz $\leq \omega/2\pi \leq$ 500 Hz below $T = 0.15|J_1|$ (Figs. S1B and S2D), with the result shown in Fig. 2E that $\alpha(T) \approx 1.2 \pm 0.1$ at the lowest temperature. Finally, the power-law $\beta$ of the SSL noise variance $\sigma^2_{\bar{\vartheta}_{x,y}}(T) \propto T^{\beta}$ below $T = 0.15|J_1|$ is predicted in Fig. 2F using $\sigma^2_{\bar{\vartheta}_{x,y}}(T) = \frac{1}{K}\sum_{k=0}^{K-1} \bar{\vartheta}^2_{x,y}(t_k,T) - \left(\frac{1}{K}\sum_{k=0}^{K-1} \bar{\vartheta}_{x,y}(t_k,T)\right)^2$. The noise variance rapidly grows down to $T = 0.15|J_1|$, then declines approximately as $T^{2.5}$. Simulation for Heisenberg spins using the same SSL Hamiltonian (Methods) predicts quantitatively distinct spin noise features and the momentum vortex is harder to identify. Here we mainly discuss the predictions for XY spins which, with the parameter set of our present simulations, conform best to the subsequently shown experimental observations in $Ca_{10}Cr_7O_{28}$. Optimization of simulation parameters in a Heisenberg-spin simulation, which might alter its correspondence with the experiment, is left for future work. To summarize, our MC simulation for XY spins using equation (1) predicts that, for $T \leq 0.15|J_1|$, the noise spectrum of a generic 2D SSL has powerful spin fluctuations at least from 1 Hz to 500 Hz, a scale-invariant power spectral density $S_{\bar{\vartheta}_{x,y}}(\omega,T) \propto \omega^{-1.2\pm0.1}$, correlation functions $C_{\bar{\vartheta}_{x,y}}(t,T) \propto -\ln t$, and a noise variance $\sigma^2_{\bar{\vartheta}_{x,y}}(T)$ diminishing approximately as $T^{2.5}$.

To explore these predictions, we perform SQUID-based flux-noise spectrometry[10,11] achieving magnetic field sensitivity approaching $\mu_0 \delta M \leq 10^{-14}$ T/$\sqrt{\text{Hz}}$ (Fig. 3A), and using cryogen-free ultra-low-vibration refrigerators in the range $10\,\text{mK} \leq T \leq 5000\,\text{mK}$ (Methods). The time-sequence of the magnetic flux $\Phi(t)$ generated by the sample magnetization $M(t) = c\Phi(t)$ within the pickup coil is measured with microsecond precision via a persistent superconducting circuit that transforms it into the flux $\Phi_{\text{S}}(t)$ at the SQUID input coil as

$$\Phi_{\text{S}}(t) = (\mathcal{M}_{\text{i}}/(L_{\text{p}} + L_{\text{i}}))\Phi(t) = (\mathcal{M}_{\text{i}}/(L_{\text{p}} + L_{\text{i}}))c^{-1}M(t) \equiv c_{\text{S}}^{-1}M(t) \quad (2)$$

Here $L_{\text{p}}$ is a pickup coil inductance, $L_{\text{i}}$ is a SQUID-input coil inductance, $\mathcal{M}_{\text{i}}$ is a mutual inductance between the SQUID and its input coil, and $c_{\text{S}}$ is a constant set by the geometry of each pickup coil (Fig. 3A). Hence, the output voltage of the SQUID $V_{\text{S}}(t)$ is related to magnetization $M(t)$ as

$$V_S(t) = g\Phi_{\text{S}}(t) = gc_{\text{S}}^{-1}M(t) \equiv a^{-1}M(t) \quad (3)$$

where $g$ is the total gain of the electronics. For a given experiment, the value of $a$ can be calibrated accurately (Methods). The time-sequences of magnetization fluctuations are recorded from $V_{\text{S}}(t)$ at each temperature $T$ as $M(t,T) = aV_{\text{S}}(t,T)$ from whence the power spectral density of magnetization noise $S_M(\omega,T) \equiv a^2 S_{V_{\text{S}}}(\omega,T)$ can be derived.

Our $Ca_{10}Cr_7O_{28}$ samples are prepared by the traveling-solvent-floating-zone method[22]. The lattice structure is confirmed by x-ray Laue diffraction, and Curie-Weiss fit $\chi = \chi_0 + \frac{C_{\text{Curie}}}{T - T_{\text{CW}}}$ of the DC magnetic susceptibility in the temperature range 50-250 K yields the Curie-Weiss temperature $T_{\text{CW}} = +2.6$ K and an effective magnetic moment $\mu_{\text{eff}} \approx 1.69\mu_{\text{B}}$ (Methods and Fig. S5B) [23]. Typical examples of the mm-scale $Ca_{10}Cr_7O_{28}$ single crystals studied are shown in Fig. S5A. The experimental setup is integrated into either a cryogen-free $^3$He refrigerator or cryogen-free $^3$He/$^4$He dilution refrigerator spanning the temperature range 10 mK $\leq T \leq$ 5000 mK. Immediately upon commencing these experiments we discovered that $Ca_{10}Cr_7O_{28}$ generates powerful magnetization noise. Figure 3B shows exemplary time-sequences of the measured magnetic flux $\Phi(t,T)$ generated by $Ca_{10}Cr_7O_{28}$ for eight selected temperatures demonstrating the intense magnetization amplitude fluctuations approaching nT amplitudes. These data are digitized by an effective 16-bit analog-to-digital converter with acquisition time interval of minimum $\Delta t = t_{k+1} - t_k = 1\ \mu$s, yielding a sequence of values $\Phi(t_k, T)$ over a continuous time epoch $\Gamma$. From this we derive the power spectral density

$$S_\Phi(\omega_j, T) \equiv \frac{1}{\pi\Gamma} \left| \Delta t \sum_{k=0}^{K-1} e^{-i\omega_j t_k} \Phi(t_k, T) \right|^2. \quad (4)$$

and consequently $S_M(\omega, T) = c^2 S_\Phi(\omega, T)$. Next we carry out a sequence of measurements consisting of varying the sample temperature from 100 mK to 800 mK in steps of 50 or 100 mK and measuring $\Phi(t_k, T)$ with $\Gamma$ =1000 s at each temperature. From that data set the $S_\Phi(\omega, T)$ for the temperature range 100 mK $\leq T \leq$ 800 mK are derived using equation (4). These $S_\Phi(\omega, T)$ spectra at $T \leq T^*$ are shown in Fig. 4A in the frequency range 0.1 Hz $\leq \omega/2\pi \leq$ 500 Hz. The full temperature range and full frequency up to 50 kHz are shown in Figs. S7A and S8. The equivalent power spectral density of magnetization noise $S_M(\omega, T)$ in units of Tesla ($M = B/\mu_0$) is shown at right. Even at this elementary stage the phenomenology of $Ca_{10}Cr_7O_{28}$ appears quite remarkable because powerful fluctuations in the spin-1/2 magnetization of a mm-scale sample are spontaneously generating magnetic fields approaching $10^{-10}$ T and occur in a frequency range 0.1 Hz $\leq \omega/2\pi \leq$ 50 kHz. Most profoundly, the $S_M(\omega, T)$ is obviously scale invariant $S_M(\omega, T) \propto \omega^{-\alpha(T)}$, and the noise power diminishes precipitously below $T^*$ (Fig. 4A).

We also measure the $Ca_{10}Cr_7O_{28}$ magnetic susceptibility $\chi(\omega, T) \equiv \frac{\mu_0 M(\omega, T)}{B(\omega)} = \chi'(\omega, T) + i\chi''(\omega, T)$ simultaneously with $S_M(\omega, T)$ (Methods). The primary superconductive coil applies homogeneous axial AC magnetic fields $B(\omega)$, whose flux does not reach the SQUID due to the balanced astatic pair of coils in the flux pickup system (Fig. 3A). If the fluctuation-dissipation theorem holds for the spin liquid (as it would not for a spin glass) then $\chi''(\omega, T)$ should equal $\pi V \mu_0 \omega S_M(\omega, T)/2k_B T$ where $V$ is the sample volume[44]. The simultaneously measured values of $\pi V \mu_0 \omega S_M(\omega, T)/2k_B T$, plotted versus $\chi''(\omega, T)$ over the range 100 mK $\leq T \leq$ 500 mK, are presented in Fig. 3C. Evidently, the fluctuation-dissipation theorem holds and dynamical equilibrium is maintained in the spin liquid state of $Ca_{10}Cr_7O_{28}$ at temperatures $T \ll T^*$ , down to at least 0.1 Hz. Another characterization technique for magnetization noise is the correlation function $C_\Phi(t, T)$ which is evaluated directly from

$$C_\Phi(t_k, T) = \frac{1}{l_{\text{ave}}} \sum_{l=0}^{l_{\text{ave}}-1} \Phi(t_l, T)\Phi(t_l + t_k, T). \tag{5}$$

The normalized correlation function $C_\Phi(t,T)/C_\Phi(t=0,T)$ is shown in Fig. 4B. As temperature is lowered, the enhanced spin correlation grows becoming $C_\Phi(t,T)/C_\Phi(t = 0,T) \sim -\ln t$ below $T^*$ (Fig. S10B). Such a logarithmic decay of correlation function is quite distinct from that of any system with a single relaxation time where $C_\Phi(t,T)/C_\Phi(t = 0,T) = \exp(-t/\tau)$, and can imply a distribution of microscopic relaxation times with probabilities[45] $p(\tau) \propto 1/\tau$.

Figure 4A presents the measured power spectral density of flux noise $S_\Phi(\omega,T)$ generated by $Ca_{10}Cr_7O_{28}$ samples for the temperature range 100 mK $\leq T \leq$ 400 mK. These $S_\Phi(\omega,T)$ data may be compared to Fig. 2C. The measured correlation function $C_\Phi(t,T)/C_\Phi(0,T)$ of flux noise generated by $Ca_{10}Cr_7O_{28}$ for 100 mK $\leq T \leq$ 800 mK in Fig. 4B may be compared with Fig. 2D. In Fig. 4C, the upper panel shows contour plots of overall power spectral density of flux noise $S_\Phi(\omega,T)$ generated by $Ca_{10}Cr_7O_{28}$ samples for the temperature range 100 mK $\leq T \leq$ 800 mK while the lower presents $C_\Phi(t,T)/C_\Phi(0,T)$ for the same range. The magnitude of power spectral density grows slowly down to $T^*$ and then diminishes rapidly below that. The coincidence of the crossover temperature $T^*$ indicates that the observed spin noise has the same origin as that of susceptibility[21], specific heat[21,25], and muon spin rotation phenomenology[21]. Figure 4D shows the measured noise power-law index $\alpha(T)$ obtained by fitting data in Fig. 4A to the function $S_\Phi(\omega,T) = A(T)\omega^{-\alpha(T)}$ in the range 0.1 Hz $\leq \omega/2\pi \leq$ 20 Hz (Fig. S7B) and this is to be compared with Fig. 2E. Finally in Fig. 4E we show the measured temperature dependence of flux noise variance $\sigma_\Phi^2(T)$ derived from Fig. 3B, showing a crossover peak at $T^*$ and diminution with power-law index $\beta \approx 2.3 \pm 0.1$; these data can be compared to the predicted temperature dependence of SSL noise variance $\sigma_\vartheta^2(T)$ in Fig. 2F. The non-monotonous rapid temperature dependence of spin noise indicates that its origin is thermal fluctuations, which is in line with the Monte Carlo simulation.

There is impressive wide-ranging agreement between Monte-Carlo simulation predictions of SSL noise phenomena in Fig. 2 and the data in Fig. 4. Firstly, magnetic field fluctuations near $10^{-10}$ T occur in a broad frequency range at least from 1 Hz $\leq \omega/2\pi \leq 50$ kHz. Secondly, the spin noise correlation function decays with a distinct form $C(t) \sim -\ln t$. Thirdly, the frequency power-law index of the spectral density $\alpha(T)$ reaches a value close to 1 at low temperatures. Finally, the magnetization noise variance grows upon cooling but then diminishes below the crossover $T^*$ with power-law index $\beta \approx 2.5$. On the other hand, the observed power spectral density characteristics are highly distinct from the $S(\omega) \propto \omega^0$ ($\alpha(T) = 0$) as predicted for $Z_2$ and $U(1)$ QSLs[8,9]. Moreover, the combined phenomenology of temperature and frequency dependence observed in $Ca_{10}Cr_7O_{28}$ spin noise is exceptional and appears inconsistent with the spin noise[5,6,8] due to quenched disorder (e.g. used to explain surface spins) or by a spin glass. Thus the quantitative correspondence between the SSL simulations and the spin noise data, including for $S_M(\omega, T)$, $C_M(t, T)$ and $\sigma_M^2(T)$ over orders of magnitude in frequency evidences the state of $Ca_{10}Cr_7O_{28}$ as a spiral spin liquid. Future studies exploring this conclusion will require extending the range of spin noise measurements to a higher frequency and performing comprehensive theoretical studies so that spin noise can be connected to the reported characteristics of neutron/muon studies. More broadly, the spin noise spectroscopy technique introduced to spin liquid studies here can indeed fingerprint a spin system, opening a completely new avenue for spin liquid research.

## FIGURES

**Figure 1 $Ca_{10}Cr_7O_{28}$ spiral spin liquid and momentum vortex**

**A.** Left: Schematic of the $Ca_{10}Cr_7O_{28}$ distorted bilayer kagome lattice. Each $Cr^{5+}$ ion hosts spin-1/2 under a tetragonal crystal field. The six $Cr^{5+}$ spin-1/2 states per unit cell occur at the sites shown. Right: Three spins on a triangular plaquette on each layer are bound by a strong ferromagnetic interaction, and they form a frustrated spin-3/2 on a monolayer honeycomb network with ferromagnetic nearest-neighbor and antiferromagnetic next-nearest-neighbor interactions[34].

**B.** Left: schematic of a spin spiral state in which the angle to the $x$-axis of the spin vector $\boldsymbol{\vartheta}$ at point $\boldsymbol{r}$ is $\theta(\boldsymbol{r})$. In a SSL, the spin spiral ground-state wavevector $\boldsymbol{Q}$ indicated by a black arrow is free to point at any in-plane angle $\Theta$. The arrangement shown here is of a unique topological defect referred to as momentum vortex, such that the line-integral on any trajectory surrounding the symmetry point is $\oint \boldsymbol{\nabla}\Theta \cdot d\boldsymbol{l} = 2\pi$. Right: contours of degenerate ground-state wavevector $\boldsymbol{Q}$ in the plane for different parameterizations of the Hamiltonian equation (1).

**C.** Schematic images of $\theta(\boldsymbol{r})$ for three simple cases. Left: a topological defect-free spin spiral state. Center: a simple momentum vortex fixed at the origin. Right: a simple momentum anti-vortex.

**Figure 2 MC simulations of spiral spin liquid noise for $Ca_{10}Cr_7O_{28}$ relevant parameters**

**A.** MC simulations of a snapshot of $\theta(\boldsymbol{r})$ on the SSL model of equation (1) and Ref. 37 using a square lattice with $N = 100 \times 100$ sites $\boldsymbol{r}$, each site with an in-plane spin unit vector $\boldsymbol{\vartheta}(\boldsymbol{r})$. Each snapshot is for a different temperature so that this sequence of SSL simulation snapshots is for approximately $T = 0.5|J_1|$, $0.1|J_1|$, $0.05|J_1|$ and $0.005|J_1|$. While the spins point random directions in a paramagnet state at high temperature, they become spatially correlated at $T = 0.1|J_1|$ corresponding to a pancake liquid state. At $T = 0.05|J_1|$, the system is in a spiral spin liquid state with spiral domains and momentum vortices, which become clearer and more rigid upon further cooling[37]. Examples of these results are demonstrated in Supplementary Movies 1 and 2.

**B.** MC predicted time sequence of average $x$-component spin $\bar{\vartheta}_x(t,T) = \frac{1}{N}\sum_{\boldsymbol{r}} \vartheta_x(\boldsymbol{r},t,T)$ at eight temperatures for $N = 40 \times 40$ sites. The average spin fluctuates at a second timescale and the amplitude of low-frequency noise grows as the system is cooled down to $T$ =0.15$|J_1|$. Below $T$ =0.15$|J_1|$ the noise amplitude gradually diminishes. We take 1 MC time step to be $\tau = 1\ \mu$s. $\bar{\vartheta}_x(t,T)$ is down sampled for visual clarity to every 500 MC steps so that time intervals shown here are $500\tau = 500\ \mu s$. The frequency component above 1 kHz is filtered out. The magnitude of magnetization noise estimated as described in Methods is indicated by the bar on the right. The simulated $\bar{\vartheta}_y(t,T)$ results are statistically equivalent as shown in Fig. S2A.

**C.** From the time sequences $\bar{\vartheta}_x(t,T)$ described in **B**, the power spectral density of simulated SSL noise $S_{\bar{\vartheta}_x}(\omega,T)$ is derived as a function of temperature $T$ and shown for seven selected temperatures. Again we take 1 MC time step to be $\tau = 1\ \mu$s. Here the error bars are the standard error of the independent MC simulation runs. The anticipated power spectral density of magnetization noise $S_M(\omega,T)$ is shown on the right-hand axis as estimated from calculations described in Methods. The spectrum shows a powerful low-frequency noise down to $\omega/2\pi =$ 1 Hz with a diminishing power below $T = 0.15|J_1|$. An example of fitted $\omega^{-\alpha}$ line (gray) is drawn and obtained $\alpha$ values are plotted in **E**.

**D.** From the time sequences $\bar{\vartheta}_x(t,T)$ described in **B**, the correlation function of simulated SSL noise $C_{\bar{\vartheta}_x}(t,T)/C_{\bar{\vartheta}_x}(0,T)$ is derived. The gray dashed line is an exemplary $-\ln t$ curve $(1 - 0.15\ln(t\ (\mathrm{ms})))$.

**E.** By fitting the power spectral density of simulated SSL noise to $S_{\bar{\vartheta}_{x,y}}(\omega,T) \propto\ \omega^{-\alpha(T)}$, the SSL noise power-law $\alpha(T)$ is derived and found to be $\alpha(T) \approx 1.2$ at the lowest temperature.

**F.** From the time sequences $\bar{\vartheta}_{x,y}(t,T)$ in **B**, the variance of simulated SSL noise $\sigma^2_{\bar{\vartheta}_{x,y}}(T)$ is presented. The variance peaks around $T = 0.15|J_1|$ and diminishes approximately as $T^{2.5}$ (blue line).

**Figure 3 Spin noise measurements in $Ca_{10}Cr_7O_{28}$**

**A.** Conceptual design of our spin noise spectrometer based on high-precision and high-bandwidth SQUID sensing of time dependent flux $\Phi(t)$ generated by the $Ca_{10}Cr_7O_{28}$ sample in the compensated superconductive pickup coil connected persistently to the SQUID input coil.

**B.** Eight typical time-sequences of the measured magnetic flux $\Phi(t)$ generated by $Ca_{10}Cr_7O_{28}$ at $100\ \mathrm{mK} \leq T \leq 800\ \mathrm{mK}$. The plotted datapoints are down sampled to every 500 $\mu s$ for visual clarity. The frequency above 1 kHz is filtered out. The equivalent spontaneous magnetization noise at the sample $\mu_0 M(t)$ is very intense, reaching almost the nT scale.

**C.** Simultaneously measured magnetization noise power spectral density $S_M(\omega, T)$ and imaginary part of AC susceptibility $\chi''(\omega, T)$ of $Ca_{10}Cr_7O_{28}$ plotted as $\pi V \mu_0 \omega S_M(\omega, T)/2k_\mathrm{B}T$ versus $\chi''(\omega, T)$ over the frequency range $0.1\ \mathrm{Hz} \leq \omega/2\pi \leq 100\ \mathrm{Hz}$ and temperature range $100\ \mathrm{mK} \leq T \leq 500\ \mathrm{mK}$. This indicates that the fluctuation-dissipation theorem $\chi''(\omega, T) = \pi V \mu_0 \omega S_M(\omega, T)/2k_\mathrm{B}T$ is predominantly valid, and that the spin liquid state in $Ca_{10}Cr_7O_{28}$ remains in dynamical equilibrium down to at least 0.1 Hz.

**Figure 4 Spiral spin liquid noise of $Ca_{10}Cr_7O_{28}$**

**A.** Typical measured power spectral density of flux noise $S_\Phi(\omega, T)$ generated by $Ca_{10}Cr_7O_{28}$ samples for the temperature range $100\ \mathrm{mK} \leq T \leq 400\ \mathrm{mK}$. Here the error bars are the standard error of separated segments (Methods). The noise spans a broad frequency range of at least 0.1 Hz $\leq \omega/2\pi \leq$ 500 Hz. The equivalent power spectral density of magnetization noise at the sample $S_M(\omega, T)$ is presented in units of Tesla on right hand axis. An example of fitted $\omega^{-\alpha}$ line (gray) is shown and obtained $\alpha$ is plotted in **D**. These data may be compared to the expected SSL spin noise spectra predicted in Fig. 2C.

**B.** Measured normalized correlation function $C_\Phi(t, T)/C_\Phi(0, T)$ of flux noise $\Phi(t)$ generated by $Ca_{10}Cr_7O_{28}$ for the temperature range $100\ \mathrm{mK} \leq T \leq 800\ \mathrm{mK}$. The gray dashed line is an exemplary $-\ln t$ curve $(1 - 0.14 \ln(t\ (\mathrm{ms})))$. This may be compared with the expected temperature dependence of SSL correlation functions predicted in Fig. 2D.

**C.** Top: overall power spectral density of flux noise $S_{\Phi}(\omega, T)$ generated by $Ca_{10}Cr_7O_{28}$ samples for the temperature range $100\ \mathrm{mK} \leq T \leq 800\ \mathrm{mK}$. The noise power is the strongest at $\omega/2\pi = 0.1$ Hz around 400 mK and gradually declines as the temperature gets away and frequency gets higher. Bottom: evolution of $C_{\Phi}(t, T)/C_{\Phi}(0, T)$ for the magnetic flux noise $\Phi(t)$ generated by $Ca_{10}Cr_7O_{28}$ at $100\ \mathrm{mK} \leq T \leq 800\ \mathrm{mK}$.

**D.** Measured noise power-law index $\alpha(T)$ obtained by fitting power spectral density $S_{\Phi}(\omega, T) = A(T)\omega^{-\alpha(T)}$ in the range $0.1\ \mathrm{Hz} \leq \omega/2\pi \leq 20\ \mathrm{Hz}$. This is to be compared with the expected temperature dependence of SSL noise power law index predicted in Fig. 2E.

**E.** Temperature dependence of measured flux noise variance $\sigma_{\Phi}^2(T)$ calculated from time-series in Fig. 3B showing a crossover peak around $T^* \sim 450$ mK. The equivalent magnetization noise variance $\sigma_M^2(T)$ is on the right axis. The noise diminishes for $T \leq 300$ mK with an approximate power law $T^{2.3}$ (blue line). These data can be compared to the expected temperature dependence of SSL noise variance $\sigma_{\vartheta}^2(T)$ as predicted in Fig. 2F.

**Acknowledgements**: We acknowledge and thank C. Carroll, J. C. Dasini, J. N. Hallen, E.-A. Kim, S. A. Kivelson, P. A. Lee, G. Luke, and S. Sondhi for key discussions and guidance. S.J.B. acknowledges support from UK Research and Innovation (UKRI) under the UK government's Horizon Europe funding guarantee [Grant No. EP/X025861/1]. J.C.S.D. and F.J. thank the MPI-CPFS for support. J.C.S.D. acknowledges support from the Moore Foundation's EPiQS Initiative through Grant GBMF9457. C.-C.H. and J.C.S.D. acknowledge support from the European Research Council (ERC) under Award DLV-788932. H. Takahashi and J.C.S.D. acknowledge support from the Royal Society under Award R64897. J.M., J.W., and J.C.S.D. acknowledge support from Science Foundation of Ireland under Award SFI 17/RP/5445. J.D.E., J.K., and J.C.S.D acknowledge support from the Moore Foundation Grant # 10599.

**Author contributions:** J.C.S.D. and H. Takagi conceived the project. H.Takahashi, P.P., M.I., Y.M., and H. Takagi synthesized and characterized the samples; C.-C.H, F.J., J.D.E., J.K., J.W., J.M. and H. Takahashi developed spin noise spectroscopy and carried out experimental measurements; H. Takahashi and S.J.B generated Monte Carlo simulations of spiral spin liquid noise, H. Takahashi developed and carried out the comprehensive analysis. J.C.S.D. and S.J.B supervised the research and wrote the paper with key contributions from C.-C.H, J.M., F.J., H. Takagi and H. Takahashi. The manuscript reflects the contributions and ideas of all authors.

# Figure 1

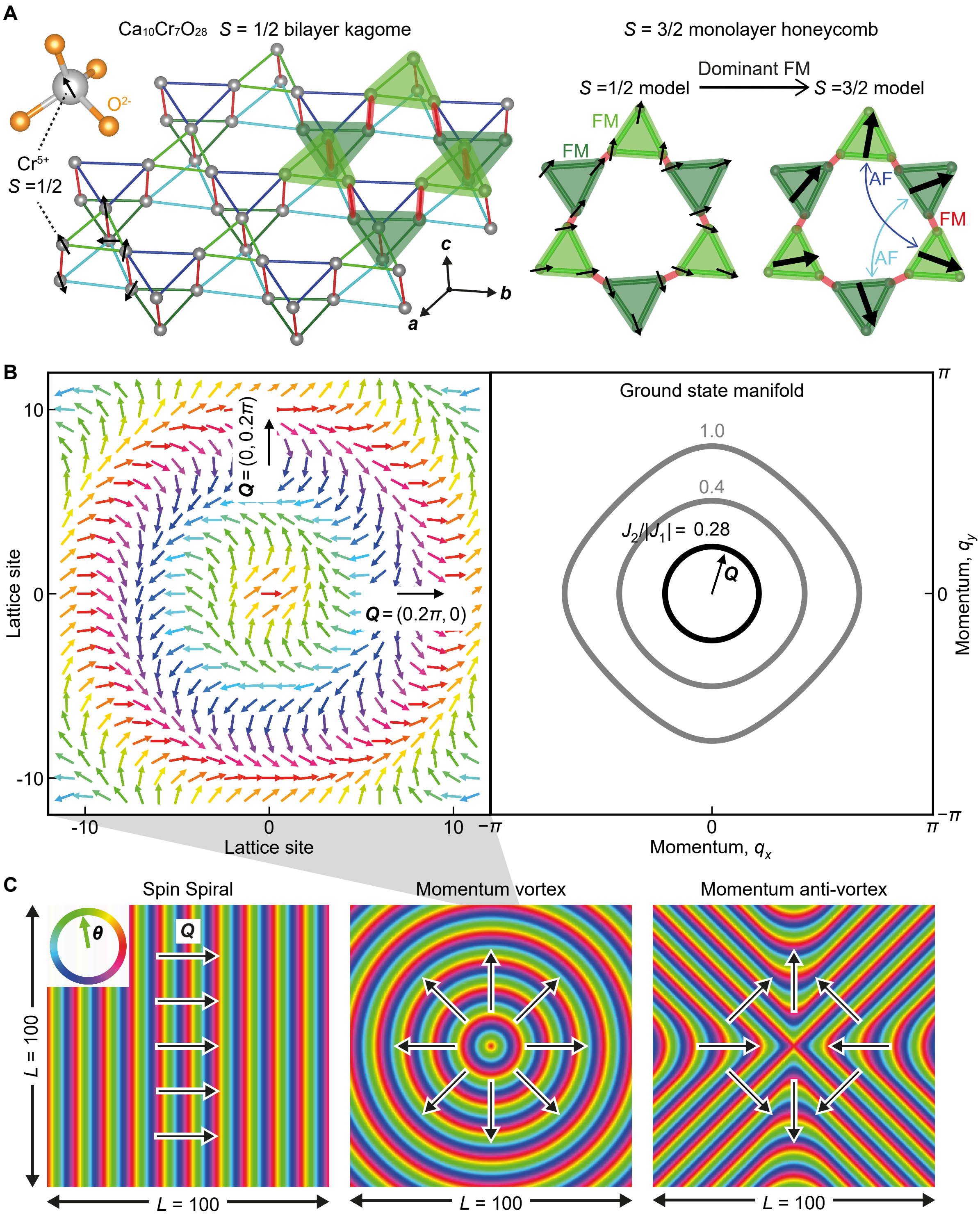
A
Ca10Cr7O28 S = 1/2 bilayer kagome
O2-
Cr5+
S =1/2
c
b
a
S = 3/2 monolayer honeycomb
S =1/2 model
Dominant FM
S =3/2 model
FM
FM
AF
FM
AF
B
Q = (0, 0.2π)
Q = (0.2π, 0)
Lattice site
Lattice site
10
0
-10
-10
0
10
Ground state manifold
1.0
0.4
J2/|J1| = 0.28
Q
Momentum, qx
Momentum, qy
−π
0
π
C
Spin Spiral
Momentum vortex
Momentum anti-vortex
θ
Q
L = 100
L = 100
L = 100
L = 100

# Figure 2

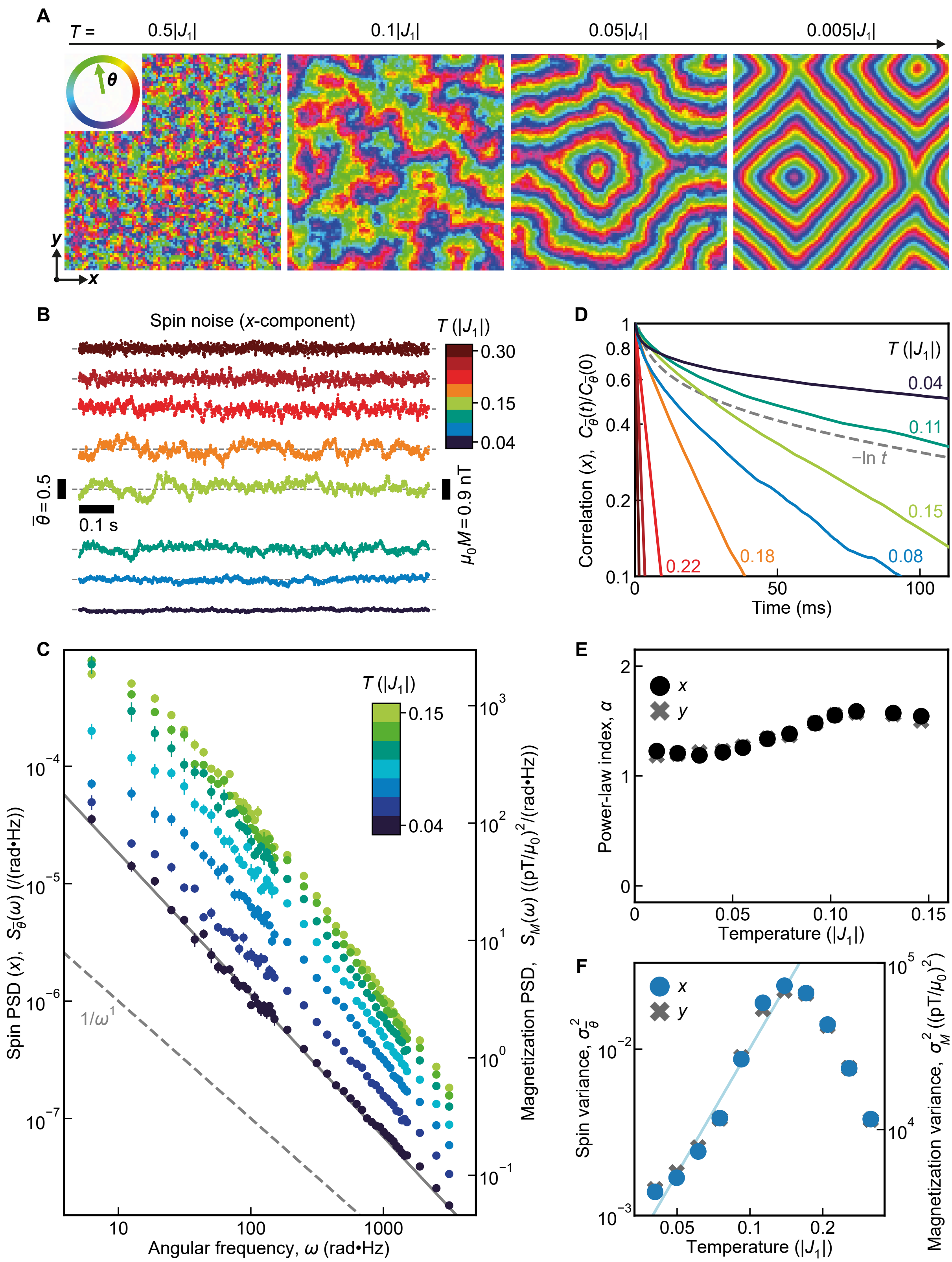
A
T =
0.5|J1|
0.1|J1|
0.05|J1|
0.005|J1|
θ
y
x
B
Spin noise (x-component)
T (|J1|)
0.30
0.15
0.04
θ̄ = 0.5
0.1 s
μ0M = 0.9 nT
C
T (|J1|)
0.15
0.04
1/ω1
Spin PSD (x), Sθ̄(ω) (/(rad•Hz))
Magnetization PSD, SM(ω) ((pT/μ0)2/(rad•Hz))
Angular frequency, ω (rad•Hz)
D
T (|J1|)
0.04
0.11
−ln t
0.15
0.22
0.18
0.08
Correlation (x), Cθ̄(t)/Cθ̄(0)
Time (ms)
E
x
y
Power-law index, α
Temperature (|J1|)
F
x
y
Spin variance, σ2θ̄
Magnetization variance, σ2M ((pT/μ0)2)
Temperature (|J1|)

# Figure 3

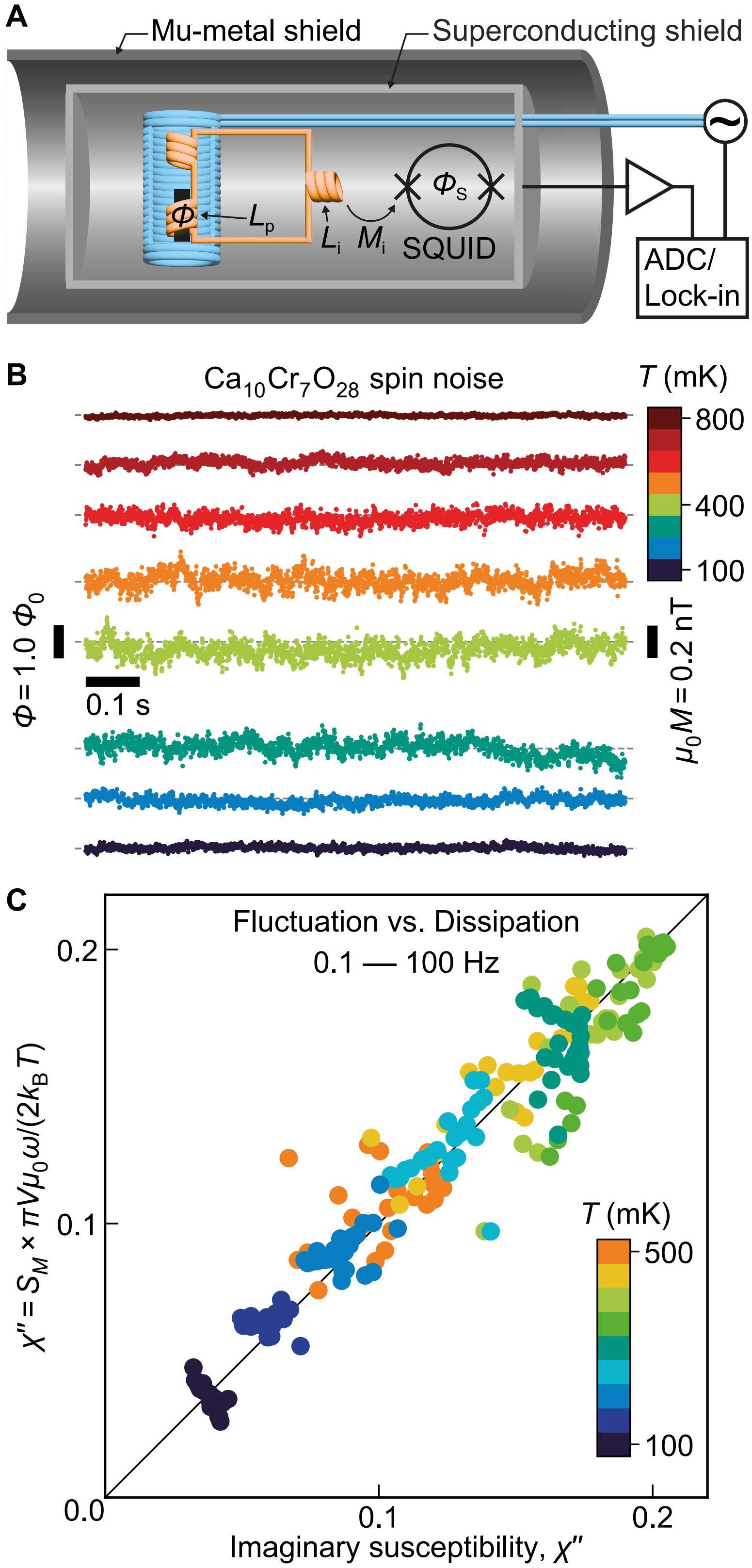
A
Mu-metal shield
Superconducting shield
Φ
L_p
L_i
M_i
Φ_S
SQUID
ADC/
Lock-in
B
Ca10Cr7O28 spin noise
T (mK)
800
400
100
Φ = 1.0 Φ0
0.1 s
μ0M = 0.2 nT
C
Fluctuation vs. Dissipation
0.1 — 100 Hz
0.2
0.1
0.0
χ'' = S_M × πVμ0ω/(2k_BT)
Imaginary susceptibility, χ''
T (mK)
500
100

# Figure 4

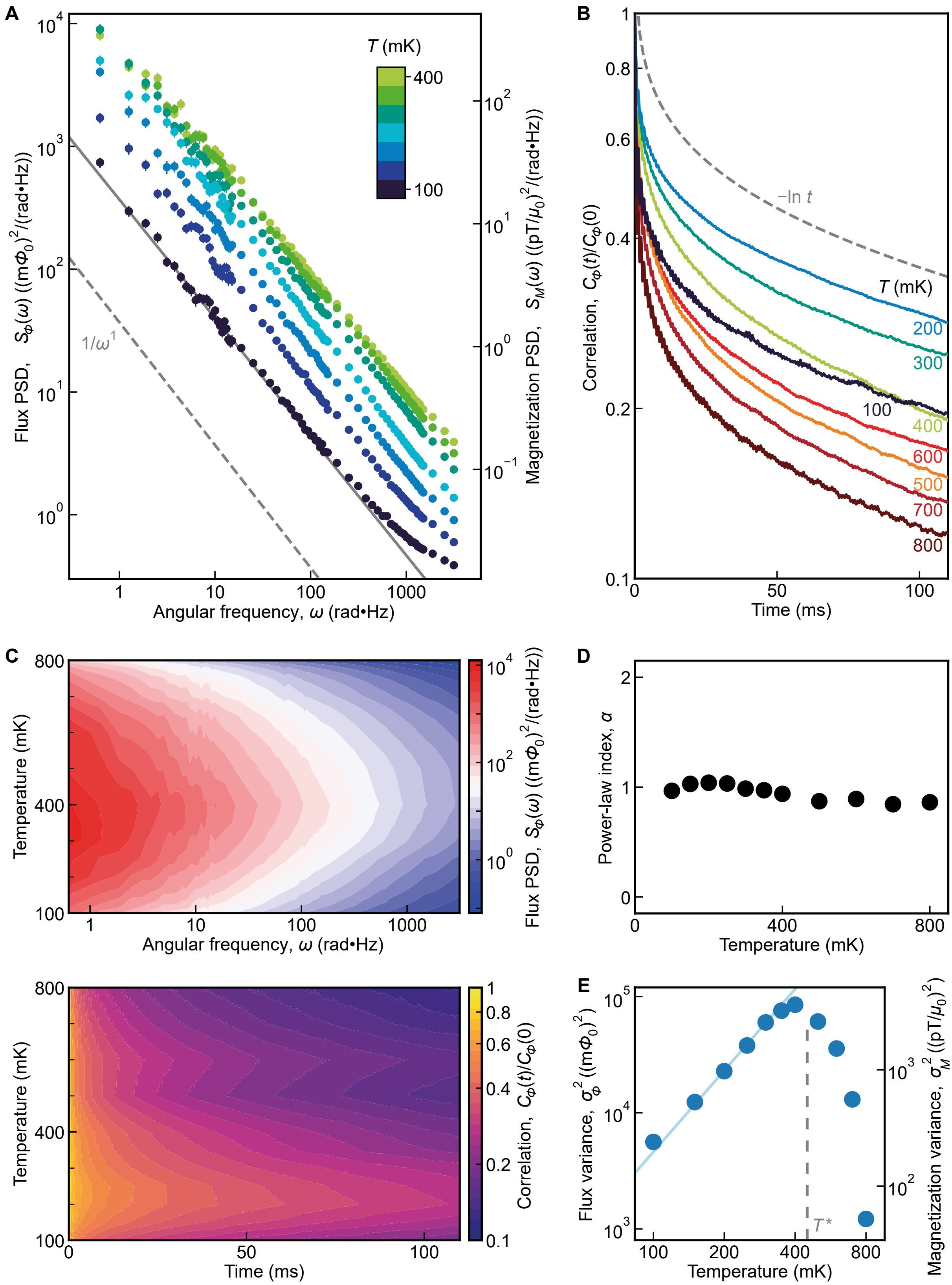

A
Flux PSD, $S_\Phi(\omega)$ ((m$\Phi_0$)$^2$/(rad•Hz))
Magnetization PSD, $S_M(\omega)$ ((pT/$\mu_0$)$^2$/(rad•Hz))
Angular frequency, $\omega$ (rad•Hz)
$T$ (mK)
400
100
$1/\omega^1$
B
Correlation, $C_\Phi(t)/C_\Phi(0)$
Time (ms)
−ln $t$
$T$ (mK)
200
300
100
400
600
500
700
800
C
Temperature (mK)
Angular frequency, $\omega$ (rad•Hz)
Flux PSD, $S_\Phi(\omega)$ ((m$\Phi_0$)$^2$/(rad•Hz))
Temperature (mK)
Time (ms)
Correlation, $C_\Phi(t)/C_\Phi(0)$
D
Power-law index, $\alpha$
Temperature (mK)
E
Flux variance, $\sigma_\Phi^2$ ((m$\Phi_0$)$^2$)
Magnetization variance, $\sigma_M^2$ ((pT/$\mu_0$)$^2$)
Temperature (mK)
$T^*$

**METHODS:**

**Quantum spin liquid noise theory**

'Fingerprinting' quantum spin liquids may, in theory, be achieved using their unique spectrum of spontaneous spin noise. For example, Ref. 8 of the main text predicts spin noise spectra of various quantum spin liquids in different parameter regimes. A separate prediction is made for $\omega \ll T$ and $\omega \gg T$ , and the $\omega \ll T$ regime is, at present, more relevant for our SQUID-based spin noise spectrometry. Another parameter controlling the noise spectrum is $d\omega/v$ , where $d$ and $v$ are the measured length scale and spinon velocity, respectively. In the $\omega \ll T$ and $d\omega/v \ll 1$ regime more relevant for our experiment, the power spectral density $S(\omega, T)$ is predicted to be frequency-independent $S(\omega, T) \propto \omega^0$ for both $Z_2$ Dirac, $Z_2$ Fermi surface, and $U(1)$ Fermi surface quantum spin liquids, with different temperature dependences[8,9].

**Simulating time evolution of spins $\boldsymbol{\theta_i(t)}$ in a spiral spin liquid model**

A classical Monte Carlo (MC) simulation, based on a generic XY model for a 2D spiral spin liquid (equation (1)) with $J_1 = -1; J_2 = 0.28; J_3 = 0.14$, was performed for $N = L \times L$ spins on a square lattice with a periodic boundary condition. The equilibrated spin configuration at each temperature was prepared by annealing from high temperature. The initial direction of the spins was randomly selected with a uniform probability. The system was cooled down from $T = 2|J_1|$ to $0.005|J_1|$ via a step-by-step equilibration at selected temperatures $T = 2 \times 0.95^r |J_1|$ where $0 \leq r \leq 117$ (exponential cooling protocol).

Following Ref. 37, two types of Monte Carlo updates were used to equilibrate the system. The first update is the standard Metropolis algorithm. A randomly selected spin attempts to flip to a new direction that is chosen with a uniform probability, and the flip is accepted with a probability of $\min(1, e^{-\Delta E/T})$ where $\Delta E$ is the change of total energy caused by the flip. The second used update is the over-relaxation update. A randomly selected spin $\boldsymbol{\vartheta}_i$ is reflected about a local exchange field $\boldsymbol{H}_i = \sum_j J_{ij} \, \boldsymbol{\vartheta}_j$, which is an energy conservation process that is empirically known to accelerate the equilibration process[46]. One-third of the Monte Carlo updates are carried out by the Metropolis update, each of which is followed by two over-relaxation updates.

One MC step consists of $N$ MC updates. $5 \times 10^4$ MC steps are performed at each temperature, amounting to a total of $6 \times 10^6$ MC steps. We performed equilibration for two system sizes $L$ = 100 and $L$ = 40. In Supplementary Movie 1 ($L$ = 100) and Movie 2 ($L$ = 40), the equilibration process from $T = 2|J_1|$ to $0.005|J_1|$ is visualized, and representative spin configuration at four temperatures are shown in Fig. 2A. The corresponding structure factor $\Sigma(\boldsymbol{q}) = \frac{1}{N}\sum_{\boldsymbol{r}_i,\boldsymbol{r}_j} e^{i\boldsymbol{q}\cdot(\boldsymbol{r}_i-\boldsymbol{r}_j)}\langle \boldsymbol{\theta}_i \cdot \boldsymbol{\theta}_j\rangle$ is shown in Fig. S11. The final spin configuration at each temperature is recorded as an equilibrated state.

Starting from the obtained equilibrated spin configuration at each temperature, we simulated the time evolution of the spins. $10^7$ MC steps consisting only of Metropolis updates are performed for the $L$ = 40 system at each temperature. We set one MC step to $\tau = 1\ \mu$s as discussed in the section 'Comparing the simulation to a realistic system', making the total time of simulation $\Gamma = 10$ s.

The time evolution of average spin $x$- and $y$-components $\bar{\vartheta}_x(t_k, T) = \frac{1}{N}\sum_i \vartheta_i^x\,(t_k, T)$ and $\bar{\vartheta}_y(t_k, T) = \frac{1}{N}\sum_i \vartheta_i^y\,(t_k, T)$ are recorded for every 10 MC steps so that the time interval of the data is $\Delta t\ = 10\tau = 10\ \mu$s. The number of data points is $K = 10^6$ with $0 \leq t_k \leq (K-1)\Delta t$. $\bar{\vartheta}_x(t_k, T)$ and $\bar{\vartheta}_y(t_k, T)$ are statistically equivalent to each other.

**Predicting physical quantities from the simulation**

We calculated the one-sided power spectral density (PSD) $S_{\bar{\vartheta}_{x,y}}(\omega_j, T)$ and correlation function $C_{\bar{\vartheta}_{x,y}}(t_k, T)$ from $\bar{\vartheta}_{x,y}(t_k, T)$.

To increase the signal-to-noise ratio of PSD, we split the total time $\Gamma$ into $P$ segments $\bar{\vartheta}_{x,y}^p(t_k, T)$ of duration $\gamma = K_p\Delta t$ $\left(\Gamma = P\gamma, 0 \leq p \leq P-1, 0 \leq t_k \leq \left(K_p - 1\right)\Delta t\right)$. The PSD is calculated for each segment.

$$S_{\bar{\vartheta}_{x,y}^p}(\omega_j, T) = \frac{1}{\pi\gamma}\left|\Delta t \sum_{k=0}^{K_p-1} e^{-i\omega_j t_k}\bar{\vartheta}_{x,y}^p(t_k, T)\right|^2, \tag{M1}$$

where $\omega_j = \frac{2\pi}{\gamma} j\ (0 \leq j \leq \frac{K_p}{2})$. PSD is obtained as the average of $P$ segments.

$$S_{\bar{\vartheta}_{x,y}}(\omega_j,T)=\frac{1}{P}\sum_{p=0}^{P-1}S_{\bar{\vartheta}^p_{x,y}}(\omega_j,T) \tag{M2}$$

Averages are further taken over 10 independent MC runs. We take as an error bar the standard error from independent runs. We used $P$ values of $10, 10^2, 10^3$ to calculate PSDs of resolution $\Delta\omega/2\pi = 1, 10, 100$ Hz. The $x$-component $S_{\bar{\vartheta}_x}(\omega_j,T)$ for $T \leq 0.15|J_1|$ and $T \leq 0.30|J_1|$ are plotted in Figs. 2C and S1A, respectively. Fig. S2B shows the $y$-component $S_{\bar{\vartheta}_y}(\omega_j,T)$ equivalent to $S_{\bar{\vartheta}_x}(\omega_j,T)$.

The PSD $S_{\bar{\vartheta}_{x,y}}(\omega_j,T)$ is fitted by a function $A(T)\omega^{-\alpha(T)}$ in the frequency range 1 Hz $\leq \omega/2\pi \leq$ 500 Hz as shown in Figs. S1B and S2D. The obtained $\alpha(T)$ is plotted in Fig. 2E.

To address the low-frequency fluctuations, measured in the experiments, fluctuations above 1 kHz are filtered out from the time sequence $\bar{\vartheta}_{x,y}(t_k,T)$. In order to do this, Fourier components $\bar{\vartheta}_{x,y}(\omega_j,T)=\Delta t\sum_{k=0}^{K-1}e^{-i\omega_j t_k}\bar{\vartheta}_{x,y}(t_k,T)$ with $\omega_j=\frac{2\pi}{\Gamma}j\left(0\leq j\leq\frac{K}{2}\right)$ are set to zero for $\omega_j/2\pi > 1$ kHz and brought back to time domain $\bar{\vartheta}'_{x,y}(t_k,T)$ by inverse Fourier transform. $\bar{\vartheta}'_{x,y}(t_k,T)$ is plotted in Figs. 2B and S2A.

The correlation function $C_{\bar{\vartheta}_{x,y}}(t_k,T)$ is calculated from $\bar{\vartheta}'_{x,y}(t_k,T)$ using the standard formula.

$$C_{\bar{\vartheta}_{x,y}}(t_k,T)=\frac{1}{l_{\text{ave}}}\sum_{l=0}^{l_{\text{ave}}-1}\bar{\vartheta}'_{x,y}(t_l,T)\,\bar{\vartheta}'_{x,y}(t_{l+k},T), \tag{M3}$$

where $l_{\text{ave}}=9\times10^5$. Averages are taken over 10 independent MC runs. The normalized correlation is calculated as $C_{\bar{\vartheta}_{x,y}}(t_k,T)/C_{\bar{\vartheta}_{x,y}}(0,T)$ and shown in Figs. 2D, S2C, and S10A.

The variance of the noise in Fig. 2F is calculated as

$$\sigma^2_{\bar{\vartheta}_{x,y}}(T)=\frac{1}{K}\sum_{k=0}^{K-1}{\bar{\vartheta}'_{x,y}}^2(t_k,T)-\left(\frac{1}{K}\sum_{k=0}^{K-1}\bar{\vartheta}'_{x,y}(t_k,T)\right)^2. \tag{M4}$$

Averages are taken over 10 independent MC runs.

## Comparing the simulation to a realistic system

The correspondence between the MC time step and the actual time is decided as follows. In Fig. S3, we show the rate of a spin flip by an angle larger than 5 degrees at each temperature.

The rate is calculated by counting the occurrence of such a spin flip in the first $5 \times 10^4$ MC steps of the time evolution of the spins in equilibrium. The rate of a spin flip larger than 5 degrees is 0.1 $(\text{MC step})^{-1}$ around $T \sim 0.15|J_1|$. The elementary local relaxation process occurs at a timescale of the order $\tau_{\text{elem}} = 10$ (MC step) at low temperature. Ref. 21 reports AC susceptibility of $Ca_{10}Cr_7O_{28}$ in the form of Cole-Cole plot in the frequency range from 100 Hz to 20 kHz. Despite the deviation at low frequency, the Cole-Cole plot is on a semicircle. This suggests a relatively sharp distribution of relaxation time at high frequency, say $\omega/2\pi$ =10 kHz corresponding to $\sim 10\ \mu$s. From this we estimate $\tau_{\text{elem}}$ to be at the order of 10 $\mu$s, namely 1 MC step = 1 $\mu$s.

Magnetization fluctuation is estimated from the average spin fluctuation. Consider $N$ spins with magnitude $s$ in a volume $V$. For large $N$, the fluctuation amplitude of average spin $\Delta\bar{\vartheta}$ and magnetization $\Delta B = \mu_0 \Delta M$ will be

$$\Delta\bar{\vartheta} \propto \frac{s\sqrt{N}}{N} = \frac{s}{\sqrt{N}}\ , \tag{M5}$$

$$\Delta B \propto \mu_0(2\mu_\text{B})\frac{s\sqrt{N}}{V} = 2\mu_0\mu_\text{B}\frac{N}{V}\Delta\bar{\vartheta}. \tag{M6}$$

In the simulation, we used $N^{\text{sim}} = 1600$ spins of $s^{\text{sim}} = 1$. This can be related to the experimental sample with $N^{\text{exp}}$ spins of $s^{\text{exp}}$ in volume $V^{\text{exp}}$.

$$\Delta\bar{\vartheta}^{\text{exp}} = \frac{s^{\text{exp}}}{s^{\text{sim}}}\frac{\sqrt{N^{\text{sim}}}}{\sqrt{N^{\text{exp}}}}\Delta\bar{\vartheta}^{\text{sim}}. \tag{M7}$$

$$\Delta B^{\text{exp}} = 2\mu_0\mu_\text{B}\frac{N^{\text{exp}}}{V^{\text{exp}}}\Delta\bar{\vartheta}^{\text{exp}} = 2\mu_0\mu_\text{B}\frac{s^{\text{exp}}}{s^{\text{sim}}}\frac{\sqrt{N^{\text{sim}}N^{\text{exp}}}}{V^{\text{exp}}}\Delta\bar{\vartheta}^{\text{sim}}. \tag{M8}$$

With $N^{\text{sim}} = 1600, s^{\text{sim}} = 1, N^{\text{exp}} = 6.2 \times 10^{18}, s^{\text{exp}} = 3/2,\ V^{\text{exp}} = 2\ \text{mm}^3$, the conversion factor to estimate $B^{\text{exp}}$ from $\bar{\vartheta}^{\text{sim}}$ is

$$2\mu_0\mu_\text{B}\frac{s^{\text{exp}}}{s^{\text{sim}}}\frac{\sqrt{N^{\text{sim}}N^{\text{exp}}}}{V^{\text{exp}}} = 1.7 \times 10^{-9}\ \text{T}. \tag{M9}$$

The scale of temperature can also be compared. By setting $J_1 = -0.15$ meV and $J_2 = 0.042$ meV, $J_2/|J_1| = 0.28$ and $J_2$ is within the two error bars of the antiferromagnetic exchange energy $0.028 \pm 0.008$ meV in the empirical Hamiltonian of $Ca_{10}Cr_7O_{28}$ [23]. This renders the variance peak temperature $T = 0.15|J_1| = 260$ mK, comparable to $\sim 400$ mK observed in the experiment.

**Classical Monte Carlo simulations in $Ca_{10}Cr_7O_{28}$**

As shown in Fig. 1A, the dominant intralayer ferromagnetic interaction in $Ca_{10}Cr_7O_{28}$ bundles up three spins on alternative plaquettes to form spin-3/2 state. Below the temperature scale set by the magnitude of these ferromagnetic interactions, a fairly accurate description of the low-energy properties can be obtained by working with effective $S$ = 3/2 spins[33]. $S$ = 3/2 magnets can be described to a good approximation in classical terms so that (semi)classical simulations[33,34] are effective. Indeed, classical Monte Carlo simulations and semiclassical simulations applied to $Ca_{10}Cr_7O_{28}$ [33,34] have very successfully reproduced the liquid-like structure factors observed in experiments.

For $Ca_{10}Cr_7O_{28}$, we turned to the simplest generic model of spiral spin liquid as has been established by Ref. 37. The simulation and the real $Ca_{10}Cr_7O_{28}$ both share the continuous contour of a spiral wave vector with an approximate $U(1)$ symmetry[21,23,25], which is the essence of the spiral spin liquid physics that renders two phases essentially the same[37], leading to the inhomogeneous spin texture of momentum vortices. This observation, together with the impressive wide-ranging agreement between the predicted SSL noise phenomenology in Fig. 2 and the measured $Ca_{10}Cr_7O_{28}$ noise data in Fig. 4, strongly indicates that our simulation captures the essence of spiral spin liquid dynamics in $Ca_{10}Cr_7O_{28}$.

In our simulation, spins are evolved via MC updates, and the dynamics due to an equation of motion is not considered. This is because the Hamiltonian (equation (1)) does not give rise to a $z$-direction exchange field that will cause precession of the XY spins. Even if a $z$-direction field existed, it would only generate a very fast periodic precession of spins at 0.1 meV ~ 10 ps that will be averaged out at the timescale of our MC simulation. We finally note that spiral spin liquid phase is found in both 2D XY spins on a square lattice[34] and 3D Heisenberg model on a honeycomb lattice[47], irrespective of the spin dimension and the underlying lattice symmetry.

**Simulating spiral spin liquid model for Heisenberg spins**

A Heisenberg Hamiltonian was initially proposed for $Ca_{10}Cr_7O_{28}$, although the possible existence of local anisotropic interactions[34,48] is discussed in more recent work. To explore the effect of spin dimension, we also simulated the spiral spin liquid model for Heisenberg spins. The system size $L = 40$, Hamiltonian equation (1), and simulation procedure are completely identical as the XY-spin case, except that the spin configuration space is changed from XY to a Heisenberg model. One independent MC run is performed, and the error bars of power spectral density and variance are calculated from the standard error of separated segments in equations (M2) and (M4).

Also in the Heisenberg system, spins get more spatially correlated at low temperatures, as seen in the representative thermalized spin configurations at four different temperatures in Fig. S4A. The power spectral density $S_{\bar{\vartheta}_x}(\omega_j, T)$ in Fig. S4B exhibits an intense low-frequency fluctuation with a power-law frequency dependence. The noise variance $\sigma^2_{\bar{\vartheta}_{x,y,z}}(T)$ in Fig. S4C drops below a peak at $T \sim 0.1\ |J_1|$, with a relatively gentle slope of approximately $T^1$. These observations indicate that the qualitative features of spiral spin liquid noise described for the XY model in the main text occur irrespective of the spin dimension. However, momentum vortices are harder to identify in the Heisenberg case, and the spin noise predictions are quantitatively distinct. With the specific parameter set of our present simulation, the XY-spin result reproduces the experimental $Ca_{10}Cr_7O_{28}$ noise quantitatively significantly better. Optimization of simulation parameters which might alter its correspondence with the experiment, especially for the Heisenberg simulation, is left for future work.

**Next developments in the simulation and theory**

At $T > T^*$, the simulated power spectral density (Figs. S1A and S2B) starts showing flat white noise, while the experimental power spectral density (Fig. S7A) retains the power law behaviour. This indicates that real $Ca_{10}Cr_7O_{28}$ has additional factors that are not fully accounted for in the simulation, such as a larger system size or additional interactions. One possible candidate could be the existence of local anisotropic interactions[34,48]. This motivates further optimization of the spiral spin liquid Hamiltonian and simulation, and most importantly, finding an analytical theory of the spin noise spectrum of dynamical spin spirals.

**Design of Noise Spectrometer/AC Susceptometer**

A $^3$He/$^4$He dilution refrigerator (Proteox MX) and a cryogen−free $^3$He refrigerator (DRY ICE 300mK TERTIA) were used to carry out our experiments.

The spectrometer on the dilution refrigerator consists of a superconducting pickup coil that is enclosed in a superconducting excitation coil and connected to the SQUID (SP550). The pickup coil is wound on a macor sample holder with an inner diameter 1.6 mm and length 10 mm. A single NbTi wire forms two in-series counter-wound 10-turn pickup coils with a total inductance $L_\mathrm{p} = 0.75\ \mu\mathrm{H}$ so that the external uniform flux is cancelled out. The SQUID input coil has an inductance of $L_\mathrm{i} = 1.74\ \mu\mathrm{H}$ and a mutual inductance to the SQUID of $1/\mathcal{M}_\mathrm{i} = 0.19\ \mu\mathrm{A}/\Phi_0$, as reported by the manufacturer (Quantum Design). The excitation coil is 10 mm long and has 101 turns of a NbTi wire. The whole circuitry is contained within two Nb cylinders covered by a mu-metal cylinder for magnetic flux shielding and is mounted on the mK-plate of the refrigerator. The first Nb cylinder is provided by the SQUID manufacturer (Quantum Design), the second Nb shield has an inner diameter of 48 mm and 2 mm in thickness, and the mu-metal cylinder has an inner diameter of 58 mm and 1 mm in thickness. The whole dilution refrigerator is built on a 6-ton table that is mechanically isolated from external vibration. To accelerate the thermalization of the sample, a 0.1 mm diameter silver wire is attached to the sample by GE varnish and the other end is thermalized to the SQUID holder. The temperature of the sample and spectrometer is measured by a Rox thermometer which is mounted on the plate close to the SQUID assembly.

The spectrometer on the $^3$He refrigerator is a superconducting pickup coil connected to the SQUID (SQ1200). 10 turns of NbTi wire with inductance $L_\mathrm{p} = 0.25\ \mu\mathrm{H}$ is wound directly around a mm-scale, bar-shaped sample and fixed with GE varnish. This pickup coil circuitry is mounted on the SQUID chip with GE varnish. The SQUID input coil has an inductance of $L_\mathrm{i} = 1.3\ \mu\mathrm{H}$ and a mutual inductance to SQUID of $1/\mathcal{M}_\mathrm{i} = 0.13\ \mu\mathrm{A}/\Phi_0$. For magnetic flux shielding, the pickup coil and SQUID are all contained within a Nb cylinder covered by a mu-metal cylinder. The spectrometer is mounted on a mechanical vibration isolator, hung under the bottom plate of the refrigerator. To accelerate thermalization of the sample, four 0.2 mm diameter brass wires are attached to the pickup circuitry by GE varnish and their other ends are in contact with a copper wire that exits the shielded region. The temperature of the sample and spectrometer is measured by a CX-1030 Cernox thermometer mounted on a vibration isolator, close to the SQUID circuitry.

**Spectrometer calibration**

The magnetic flux picked up from the sample $\Phi(t)$ and the output voltage of the SQUID $V_\mathrm{S}(t)$ are related by a simple constant

$$\Phi(t) = \frac{L_\mathrm{p} + L_\mathrm{i}}{\mathcal{M}_\mathrm{i}} \frac{1}{g} V_\mathrm{S}(t). \quad \text{(M10)}$$

$g = g_\mathrm{SQUID} g_\mathrm{preamp}$ consists of the conversion factor from flux to voltage in the SQUID $g_\mathrm{SQUID}$ and subsequent gain from the preamplifier $g_\mathrm{preamp}$. $g_\mathrm{SQUID}$ is determined from the voltage jump due to a $\Phi_0$-flux jump of the SQUID. For the highest sensitivity setting, $g_\mathrm{SQUID} = 0.73\ \mathrm{V}/\Phi_0$ for the SP550 SQUID and $g_\mathrm{SQUID} = 9.85\ \mathrm{V}/\Phi_0$ for the SQ1200 SQUID. Using these values, $V_\mathrm{S}(t)$ is converted to $\Phi(t)$.

$\Phi(t)$ is further converted to the magnetization $\mu_0 M(t)$ with the relation

$$\mu_0 M(t) = \frac{1}{NA} \Phi(t), \quad \text{(M11)}$$

where $N$ is the number of turns of the pickup coil and $A$ is the area of sample cross section. $N = 10$ for all setups. $A = 1\ \mathrm{mm}^2$ for Sample 1 in the dilution refrigerator.

**$Ca_{10}Cr_7O_{28}$ Sample Preparation**

As described in Ref. 22, $Ca_{10}Cr_7O_{28}$ crystals are synthesized in a two-step process, including a solid-state reaction of $Ca_{10}Cr_7O_{28}$ powder and a travelling-solvent-floating-zone method for the single crystal growth. First, powder of $CaCO_3$ and $Cr_2O_3$ was mixed with a molar ratio of 3:1, sintered at 1000 °C for 24 hours, and rapidly quenched to room temperature. The sintering process was repeated after grinding and the addition of $Cr_2O_3$ powder until phase pure powder of $Ca_{10}Cr_7O_{28}$ was obtained. This powder was packed in a rod that is sintered at 1020 °C for 12 hours followed by a rapid quench to room temperature. This rod was used as a feed rod of the floating-zone growth, while a solvent was separately prepared following the same procedure from the powder of $CaCO_3$ and $Cr_2O_3$ with a molar ratio of 5:2. The growth was carried out in a 0.22 MPa oxygen pressure at 1 mm/hr using an optical floating zone furnace. The resulting single crystal was washed with HCl and then with $H_2O$. X-ray diffraction on a ground small piece confirms the phase purity. Fig. S5B shows the DC susceptibility of a typical single crystal measured in MPMS (Quantum Design). Fitting by $\chi = \chi_0 + \frac{C_{\mathrm{Curie}}}{T - T_{\mathrm{CW}}}$ in the temperature range $50\ \mathrm{K} \leq T \leq 250\ \mathrm{K}$ yields $T_{\mathrm{CW}} = +2.6\ \mathrm{K}$ and an effective magnetic moment $\mu_{\mathrm{eff}} \approx 1.69\mu_{\mathrm{B}}$ that are comparable to the existing literature[23].

Photos of the three $Ca_{10}Cr_7O_{28}$ samples are shown in Fig. S5A. The long direction of the bar is identified to be the c-axis. Sample 1' and Sample 2 are from the same growth while Sample 3 is from a separate growth. Sample 1 is obtained by later polishing down Sample 1' to fit into the spectrometer of our dilution refrigerator. The measured magnetic noise from these three samples are consistent with each other, as described in the section 'Repeatability of spin noise spectrum in different $Ca_{10}Cr_7O_{28}$ samples'.

**Noise measurement**

The results in the main text were measured in both the dilution refrigerator and the $^3$He refrigerator. In the dilution refrigerator, the temperature of $Ca_{10}Cr_7O_{28}$ Sample 1 and the spectrometer was controlled by heaters from 100 mK to 500 mK in steps of 50 mK with the temperature stability of 1 mK. In the $^3$He refrigerator, the temperature of the Sample 2 and the spectrometer was controlled by heaters from 300 mK to 800 mK in steps of 100 mK with the stability of 1 mK. In both setups, samples were thermalized for at least 15 minutes after the thermometer reading got stabilized at the target temperature. The overall circuit diagram for the noise measurement is shown in Fig. S6A. The output voltage of the SQUID $V_{\mathrm{S}}(t)$ was recorded by an effective 16-bit ADC (Moku:Pro) for 1000 s at a sampling rate of 20 kSa (a time interval of 50 $\mu$s). Between the SQUID and the ADC, a preamplifier (SR560) was used to amplify the SQUID output signal by an appropriate gain and to apply a 0.03 Hz 6 dB/Oct high-pass and 30 kHz 6 dB/Oct low-pass filter. For the SQUID output of the $^3$He refrigerator setup, there was further filtering by a 5 kHz, 4-pole low-pass filter. With an identical setup, the SQUID background noise was measured for a nonmagnetic nylon sample at 800 mK.

To confirm the reproducibility, the noise signal of Sample 1', Sample 2, and Sample 3 was measured in the $^3$He refrigerator. At temperatures from 275 mK to 800 mK in steps of 25 mK, the output voltage was recorded for 100 s at 1 MSa (a time interval of 1 $\mu$s) and 1.6

Hz AC coupling filter at the ADC input was used. A nonmagnetic nylon block of a comparable size was measured in the same condition in temperature steps of 100 mK.

An extended-bandwidth measurement was performed for Sample 2 in the $^3$He refrigerator. The SQUID sensitivity was changed to medium $g_{\mathrm{SQUID}} = 0.985\ \mathrm{V}/\Phi_0$ and the frequency cutoff of the SQUID was extended to $f_{3\mathrm{dB}} \sim 300$ kHz by changing the internal capacitor of the SQUID feedback loop. At temperatures from 300 mK to 800 mK in steps of 100 mK, the output voltage of SQUID $V_\mathrm{S}(t)$ was recorded for 100 s at 1 MSa. The SQUID background noise was measured for a nonmagnetic nylon sample at 275 mK for 10 s at 1 MSa.

**Noise analysis**

In the main text, one-sided power spectral density (PSD) $S_\Phi(\omega_j, T)$ and correlation function $C_\Phi(t_k, T)$ are calculated from the experimental noise data $\Phi(t_k, T)$.

To increase the signal-to-noise ratio of the PSD, $\Phi(t_k, T)$ with a total time $\Gamma$ =1000 s and a time interval $\Delta t = 50\ \mu$s is split into $P$ segments $\Phi^p(t_k, T)$ of duration $\gamma = K_p \Delta t$ $\left(\Gamma = P\gamma, 0 \le p \le P-1, 0 \le t_k \le (K_p - 1)\Delta t\right)$. The PSD for each segment is calculated from

$$S_{\Phi^p}(\omega_j, T) = \frac{1}{\pi\gamma}\left|\Delta t \sum_{k=0}^{K_p-1} e^{-i\omega_j t_k}\Phi^p(t_k, T)\right|^2, \tag{M12}$$

where $\omega_j = \frac{2\pi}{\gamma} j$ $(0 \le j \le \frac{K_p}{2})$. In a next step, the PSD is obtained as the average of $P$ segments.

$$S_\Phi(\omega_j, T) = \frac{1}{P}\sum_{p=0}^{P-1} S_{\Phi^p}(\omega_j, T), \tag{M13}$$

with the standard error used as an error bar. $P = 10^2, 10^3, 10^4, 10^5$ was used to calculate the PSD of resolution $\Delta\omega/2\pi = 10^{-1}, 10^0, 10^1, 10^2$ Hz. The resulting PSD is plotted in Fig. 4A for 100 mK $\le T \le$ 400 mK and in Fig. S7A for 100 mK $\le T \le$ 800 mK.

In Fig. S7B, the PSD $S_\Phi(\omega_j, T)$ is fitted by a function $A(T)\omega^{-\alpha(T)}$ in the frequency range 0.1 Hz $\le \omega/2\pi \le$ 20 Hz. The obtained $\alpha(T)$ is plotted in Fig. 4D.

To get rid of the partial contribution of the electronic noise at high frequency and the slow temperature fluctuation at low frequency, the fluctuation below 0.05 Hz and that above 1 kHz are filtered out from the time sequence $\Phi(t_k, T)$. The filtering was done by setting the Fourier components $\Phi(\omega_j, T) = \Delta t \sum_{k=0}^{K-1} e^{-i\omega_j t_k} \Phi(t_k, T)$ below 0.05 Hz and above 1 kHz to zero, and then transforming the remaining signal $\Phi(\omega_j, T)$ back into the time domain $\Phi'(t_k, T)$ using an inverse Fourier transform. The filtered data is shown in Fig. 3B.

The correlation function $C_\Phi(t_k, T)$ is calculated from $\Phi'(t_k, T)$ using the standard formula

$$C_\Phi(t_k, T) = \frac{1}{l_{\text{ave}}} \sum_{l=0}^{l_{\text{ave}}-1} \Phi'(t_l, T)\, \Phi'(t_{l+k}, T)\,, \tag{M14}$$

where $l_{\text{ave}} = 1.9 \times 10^7$. The normalized correlation is calculated as $C_\Phi(t_k, T)/C_\Phi(0, T)$ and shown in Fig. 4B and Fig. S10B.

The variance of the noise, shown in Fig. 4E, is calculated as

$$\sigma_\Phi^2(T) = \frac{1}{K}\sum_{k=0}^{K-1} \Phi'^2(t_k, T) - \left(\frac{1}{K}\sum_{k=0}^{K-1} \Phi'(t_k, T)\right)^2 . \tag{M15}$$

The fitting by $\sigma_\Phi^2(T) \propto T^\beta$ is performed in the temperature range of 100 mK $\leq T \leq$ 300 mK to give $\beta = 2.3 \pm 0.1$.

The noise data of Sample 1', Sample 2, and Sample 3, measured with a total time $\Gamma$ =100 s and a time interval $\Delta t = 1\ \mu$s, is used to calculate the PSD with $P = 10^3$ that is shown in Fig. S5C. The variance is calculated after filtering the fluctuation below 0.05 Hz and above 10 kHz, as shown in Fig. S5D.

In Fig. S8, we show the PSD of Sample 2 and Nylon in the full frequency range from 0.1 Hz to 50 kHz. The frequency range $\omega/2\pi$ below 3 kHz is calculated from the noise data with $\Gamma$ =1000 s and $\Delta t = 50\ \mu$s using $P = 10^2, 10^3, 10^4, 10^5$. The frequency range above 3 kHz is calculated from the extended-frequency measurement data of Sample 2 (Nylon) with $\Gamma = 100$ s (10 s) and $\Delta t = 1\ \mu$s using $P = 10^5, 10^6$ ($10^4, 10^5$). The noise of Nylon corresponding to the background noise level is enhanced by ~10 times for the extended-frequency measurement. In Fig. S8, the PSD of Sample 2 above 3 kHz is plotted after subtraction of the Nylon noise. The PSD of Sample 2 below and above 3 kHz smoothly connects to each other.

**Noise data from different spectrometers**

The experimental noise data $\Phi(t,T)$ in Fig. 3B was measured in two different samples Sample 1 and Sample 2 for different temperature ranges. The amplitude of magnetization noise $B(t)$ is dependent on a measured sample volume, and the amplitude of $\Phi(t)$ is further dependent on the area of the sample cross section. Thus, a scale factor is naturally required to patch up the results for two samples with different geometry. In principle, one can only match the scale of either $\Phi(t)$ or $B(t)$. Throughout the paper, the matching of $\Phi(t)$ is prioritized, and $B(t)$ is converted from $\Phi(t)$ using the geometry of Sample 1 for both samples.

Fig. S9 shows the comparison of power spectral density from the two samples at an overlapping temperature 300 mK. Here, the $\Phi(t)$ of Sample 2 is scaled from an original value by a factor of 0.95 (i.e. $S_{\Phi}(\omega)$ by $0.95^2$) so that two data coincide.

**Repeatability of spin noise spectrum in different $Ca_{10}Cr_7O_{28}$ samples**

The noise measurement was repeated in $^3$He refrigerator for three $Ca_{10}Cr_7O_{28}$ samples and a nonmagnetic nylon block of a comparable size shown in Fig. S5A. In Fig. S5C, we show the power spectral density of each sample at temperatures of 300 mK, 500 mK, 700 mK, and 800 mK. All $Ca_{10}Cr_7O_{28}$ samples have a scale-invariant power spectral density. The nylon noise floor barely changes over temperature and remains much smaller than the signal of $Ca_{10}Cr_7O_{28}$. Fig. S5D shows the temperature dependence of variance. All $Ca_{10}Cr_7O_{28}$ samples show a peak around $T \sim 400$ mK. Thus, the reported behavior of $Ca_{10}Cr_7O_{28}$ noise is robust.

**AC susceptibility measurement**

We performed the AC susceptibility measurement of Sample 1 in the dilution refrigerator for temperatures from 100 mK to 500 mK in 50 mK steps with the stability of 1 mK. At each target temperature, the sample was thermalized for 20 minutes after the thermometer reading got stabilized. The overall circuit diagram for the AC susceptibility measurement is shown in Fig. S6B. An AC magnetic field with a root-mean-square magnitude of $B_{\text{exc}} = 60$ nT in the frequency range 0.1 Hz $\leq \omega/2\pi \leq$ 101 Hz was applied using the reference output of the lock-in amplifier (SR830) in series of a 20 kΩ resistor and the excitation coil circuitry.

The output voltage of the SQUID at the lowest sensitivity $g_{\mathrm{SQUID}} = 7.3\ \mathrm{mV}/\Phi_0$ was fed into the lock-in amplifier to measure the in-phase and out-of-phase components. At each frequency, 10 measurements were performed and the results were averaged. The zero phase was set by performing a calibration experiment with a superconducting indium wire of the size comparable to the sample. The time constant of the lock-in amplifier was set at 30 s, 10 s, 3 s, 300 ms for frequencies 0.1-0.3 Hz, 0.5-0.9 Hz, 1-11 Hz, and 21-101 Hz, respectively. The low-pass filter was set at 18 dB/oct for all frequencies. The sensitivity of the lock-in amplifier was set to an appropriate value between 10 mV and 50 mV for different temperatures.

**AC susceptibility analysis**

The out-of-phase output of the lock-in amplifier $V_Y(\omega)$ was converted to magnetization $M_Y(\omega)$ using equation (3) and then to the imaginary susceptibility by

$$\chi''(\omega) = -\frac{\mu_0 M_Y(\omega)}{B_{\mathrm{exc}}}. \tag{M16}$$

The fluctuation-dissipation theorem relates the imaginary part of the susceptibility $\chi''(\omega, T)$ with the one-sided power spectral density of magnetization noise $S_M(\omega, T)$ as

$$\chi''(\omega, T) = \mu_0 V \frac{\pi \omega S_M(\omega, T)}{2 k_{\mathrm{B}} T}, \tag{M17}$$

where $T$ is the temperature and $V$ is the measured volume of the sample that is set to 2 $\mathrm{mm}^3$ here[44]. In Fig. 3C, the right-hand side is plotted against the left-hand side. 11-101 Hz data of $\chi''(\omega, T)$ is matched to 10-100 Hz data of $S_M(\omega, T)$ with $\Delta\omega/2\pi = 10$ Hz.

**Spin noise and AC susceptibility**

Spin noise power spectral density $S_M(\omega, T)$ and imaginary part of AC magnetic susceptibility $\chi''(\omega, T)$ are related by the fluctuation-dissipation theorem (FDT) in an equilibrium state

$$S_M(\omega, T) = \frac{1}{V \mu_0} \frac{2 k_{\mathrm{B}} T}{\pi \omega} \chi''(\omega, T). \tag{M18}$$

When FDT is valid, both measurements should yield the same quantities, although there is a key difference that spin noise measurement does not require the application of any magnetic field. This is beneficial in studies of spin liquids where a small field could alter an underlying spin state, and also a simpler design of the spectrometer makes it a promising approach to performing local measurements and extending the measurement frequency.

When the relaxational time far exceeds the experimental time, for example in a spin glass state, the FDT is expected to break down[44]. We recently exemplified this situation in $Dy_2Ti_2O_7$ [14]. Such FDT breaking is not observed in $Ca_{10}Cr_7O_{28}$ at least down to 100 mK and 0.1 Hz.

Literature searches do not find reports of spin noise measurement (i.e. magnetization noise measurement) in well-known spin glasses CuMn or AuFe, although they do exist in some spin glass compounds such as $(AI_2O_3)_{0.1}(MnO)_{0.5}(SiO_2)_{0.4}$ [49] and $Eu_{0.4}Sr_{0.6}S$ [50]. These papers support the validity of FDT, but only based on a few high-frequency points well above 1 Hz. A comprehensive study of FDT in these conventional spin glasses is motivated.

## Spin noise and charge noise

Spin noise spectroscopy can be compared to the charge noise spectroscopy. SQUID spin noise spectroscopy measures the fluctuation of magnetization $M(t)$, and is controlled by the spin precession and relaxation. On the other hand, charge noise spectroscopy measures the fluctuation of current noise density $J(t)$, which corresponds to the oscillation and scattering of charged particles momentum.

By defining an electric susceptibility $P = \chi_e D$, where $P$ is the polarization and $D$ is the electric displacement field, the polarization noise power spectral density is expressed as

$$S_P(\omega, T) = \frac{\varepsilon_0}{V}\frac{2k_\mathrm{B}T}{\pi\omega}\chi_e''(\omega, T), \tag{M19}$$

where $\varepsilon_0$ is vacuum permittivity. This is an electric version of Equation (M18).

In the similar manner, the noise of current density $J$ can be expressed by an AC conductivity. By defining the AC conductivity as $J = (\sigma/\varepsilon_0)D$, the current density power spectral density is

$$S_J(\omega, T) = \frac{1}{V}\frac{2k_\mathrm{B}T}{\pi}\sigma'(\omega, T). \tag{M20}$$

Equations (M19) and (M20) would be most relevant for dielectric material. We note that SQUID spin noise spectrometer is also sensitive to charge noise from a metallic compound, but in a more complex fashion because here the current generates magnetic field through the Biot-Savert law.

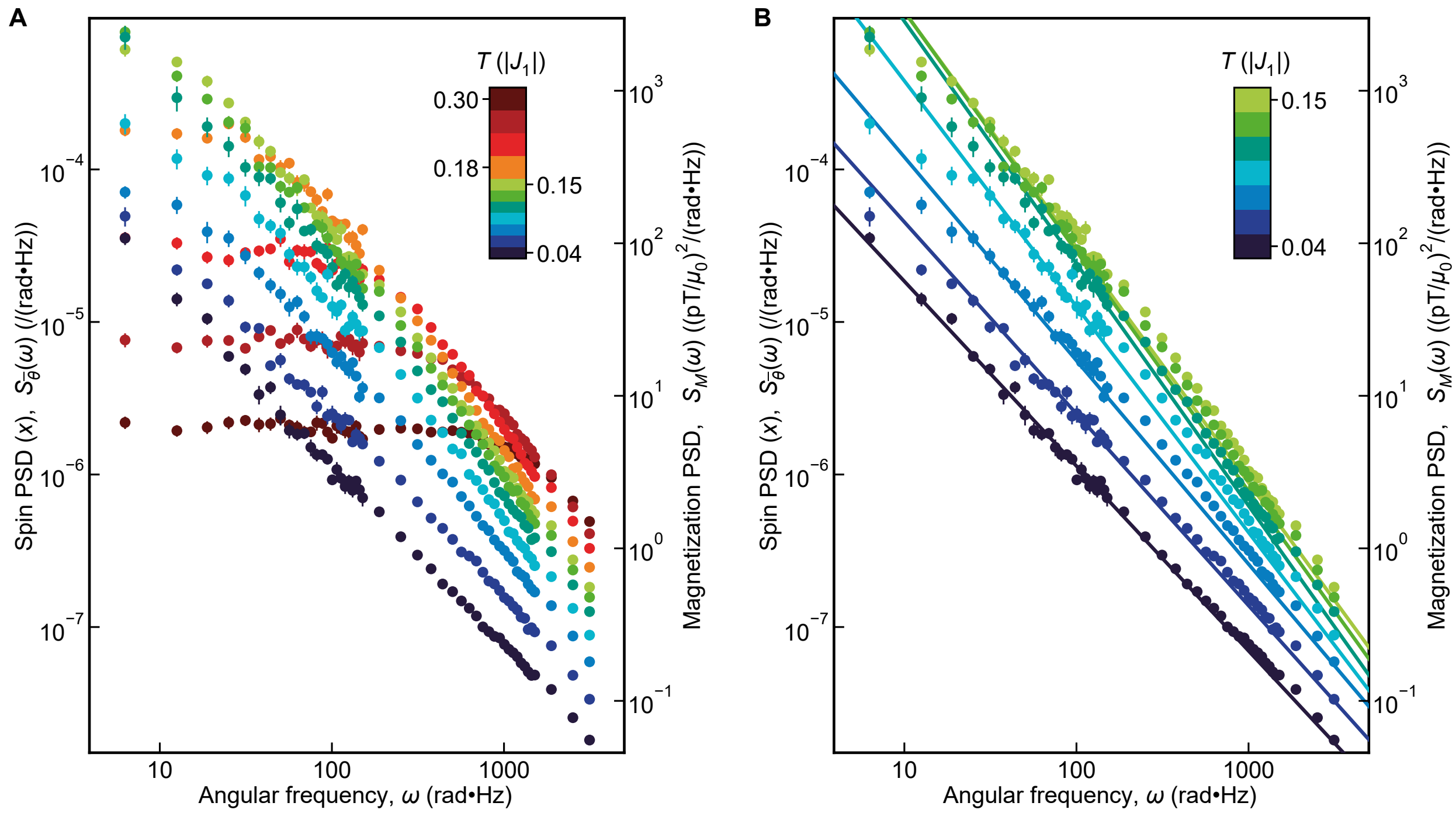

**Fig. S1. The simulated spin noise power spectral density $S_{\bar{\vartheta}_x}(\omega, T)$ fit by $A(T)\omega^{-\alpha(T)}$.**

**A.** The power spectral density of simulated spiral spin liquid noise $S_{\bar{\vartheta}_x}(\omega, T)$ for eleven selected temperatures in the range of $0.04|J_1| \leq T \leq 0.30|J_1|$. 1 MC time step is set at $\tau = 1\,\mu s$. Error bars are the standard error of the independent MC simulation runs. The anticipated power spectral density of magnetization noise $S_M(\omega, T)$ is shown on the right-hand axis as estimated from calculations described in this document.

**B.** Fitting of the simulated power spectral density $S_{\bar{\vartheta}_x}(\omega, T) = A(T)\omega^{-\alpha(T)}$ in the range of $1\text{ Hz} \leq \omega/2\pi \leq 500\text{ Hz}$ for seven selected temperatures in the range of $0.04|J_1| \leq T \leq 0.15|J_1|$.

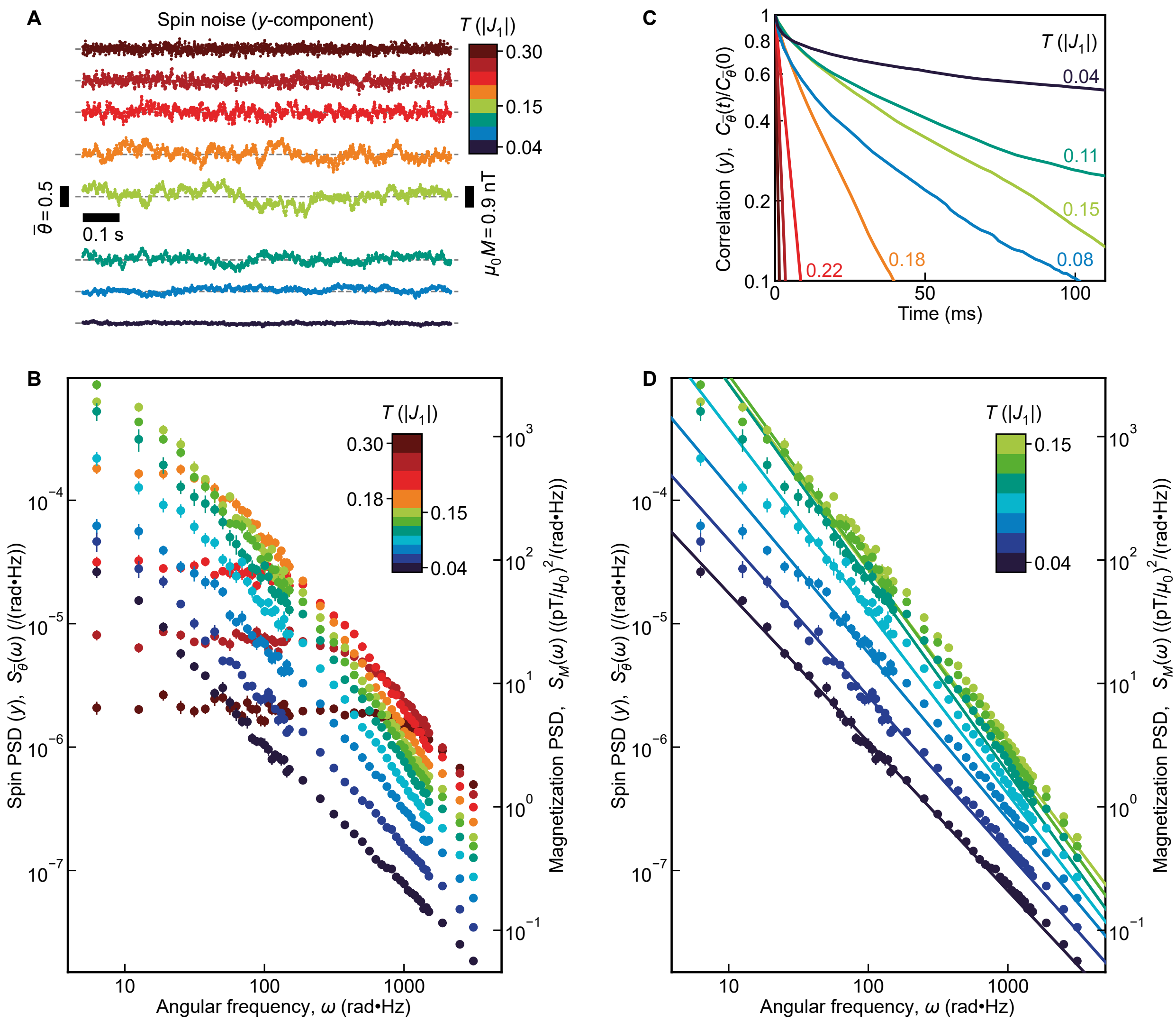

**Fig. S2. $y$-component of average spin from MC simulations of spiral spin liquid noise.**

**A.** MC-predicted time sequence of average $y$-component spin $\bar{\vartheta}_y(t,T) = \frac{1}{N}\sum_{\boldsymbol{r}} \vartheta_y(\boldsymbol{r},t,T)$ at eight temperatures for $N = 40 \times 40$ sites, equivalent to $\bar{\vartheta}_x(t,T)$ in Fig. 2B. We take 1 MC time step to be $\tau = 1\ \mu s$. $\bar{\vartheta}_y(t,T)$ is down sampled for visual clarity to every 500 MC steps so that time intervals shown are $500\tau = 500\ \mu s$. The frequency component above 1 kHz is filtered out.

**B.** The power spectral density of simulated average $y$-component spin $S_{\bar{\vartheta}_y}(\omega,T)$ for the eleven selected temperatures, comparable to $S_{\bar{\vartheta}_x}(\omega,T)$ in Figs. 2C and S1A. 1 MC time step is $\tau = 1\ \mu s$. Error bars are the standard error of the independent MC simulation runs. The anticipated power spectral density of magnetization noise $S_M(\omega, T)$ is shown on the right-hand axis as estimated from calculations described in this document.

**C**. The correlation function of average $y$-component spin $C_{\bar{\vartheta}_y}(t, T)/C_{\bar{\vartheta}_y}(0, T)$ comparable to $C_{\bar{\vartheta}_x}(t, T)/C_{\bar{\vartheta}_x}(0, T)$ in Fig. 2D.

**D**. Fitting of the simulated power spectral density $S_{\bar{\vartheta}_y}(\omega, T) = A(T)\omega^{-\alpha(T)}$ in the range of $1\ \text{Hz} \le \omega/2\pi \le 500\ \text{Hz}$ for seven selected temperatures in the range of $0.04|J_1| \le T \le 0.15|J_1|$.

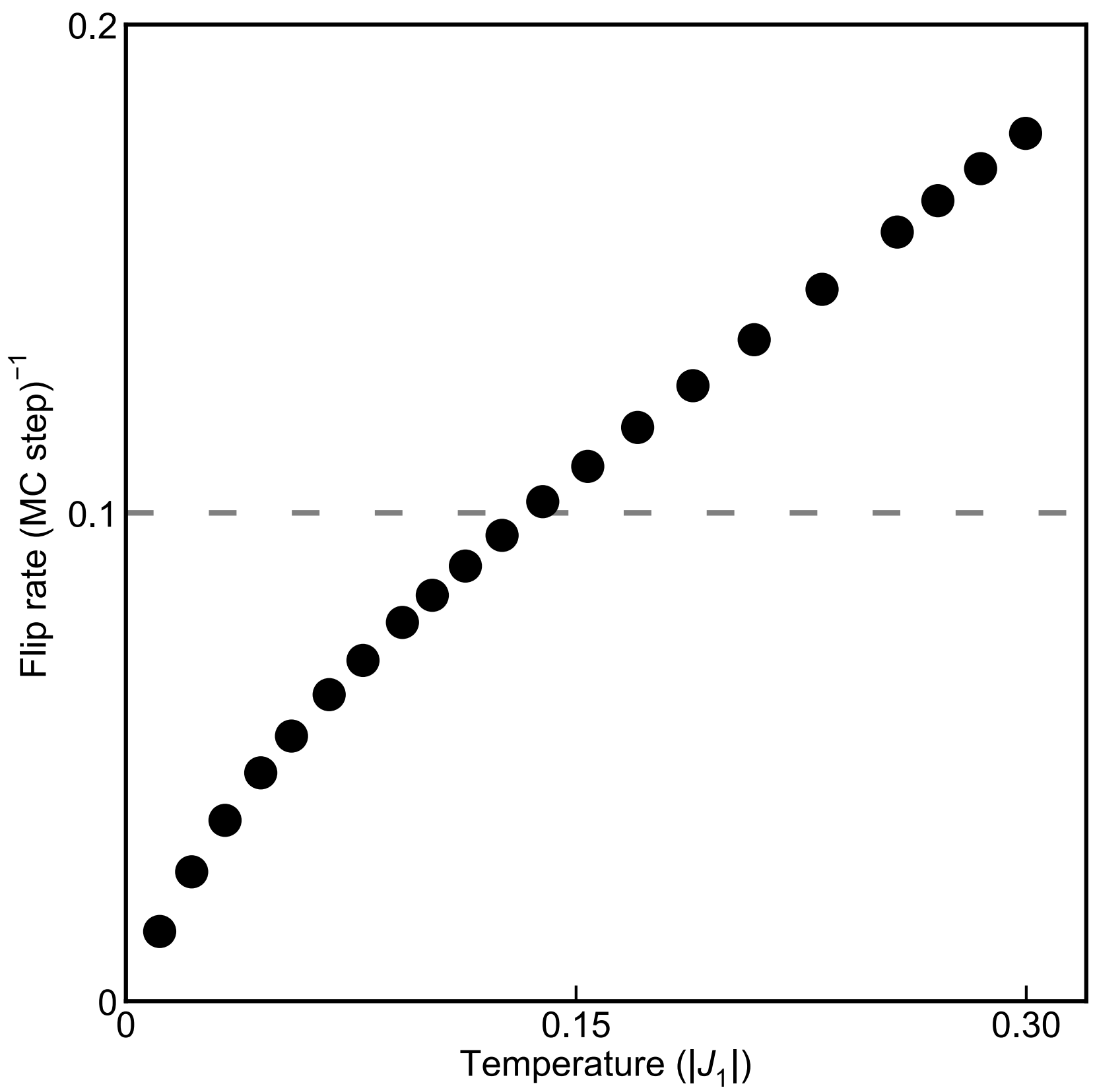

**Fig. S3. The rate of a spin flip by an angle larger than 5 degrees in the Monte Carlo simulation.**

The rate of a spin flip by an angle larger than 5 degrees at each temperature counted in the first $5 \times 10^4$ MC steps of the time evolution of $N = 40 \times 40$ spins in equilibrium. The rate is 0.1 (MC step)$^{-1}$ around $T \sim 0.15|J_1|$, suggesting an elementary local relaxation process timescale of the order $\tau_{\text{elem}} = 10$ (MC step).

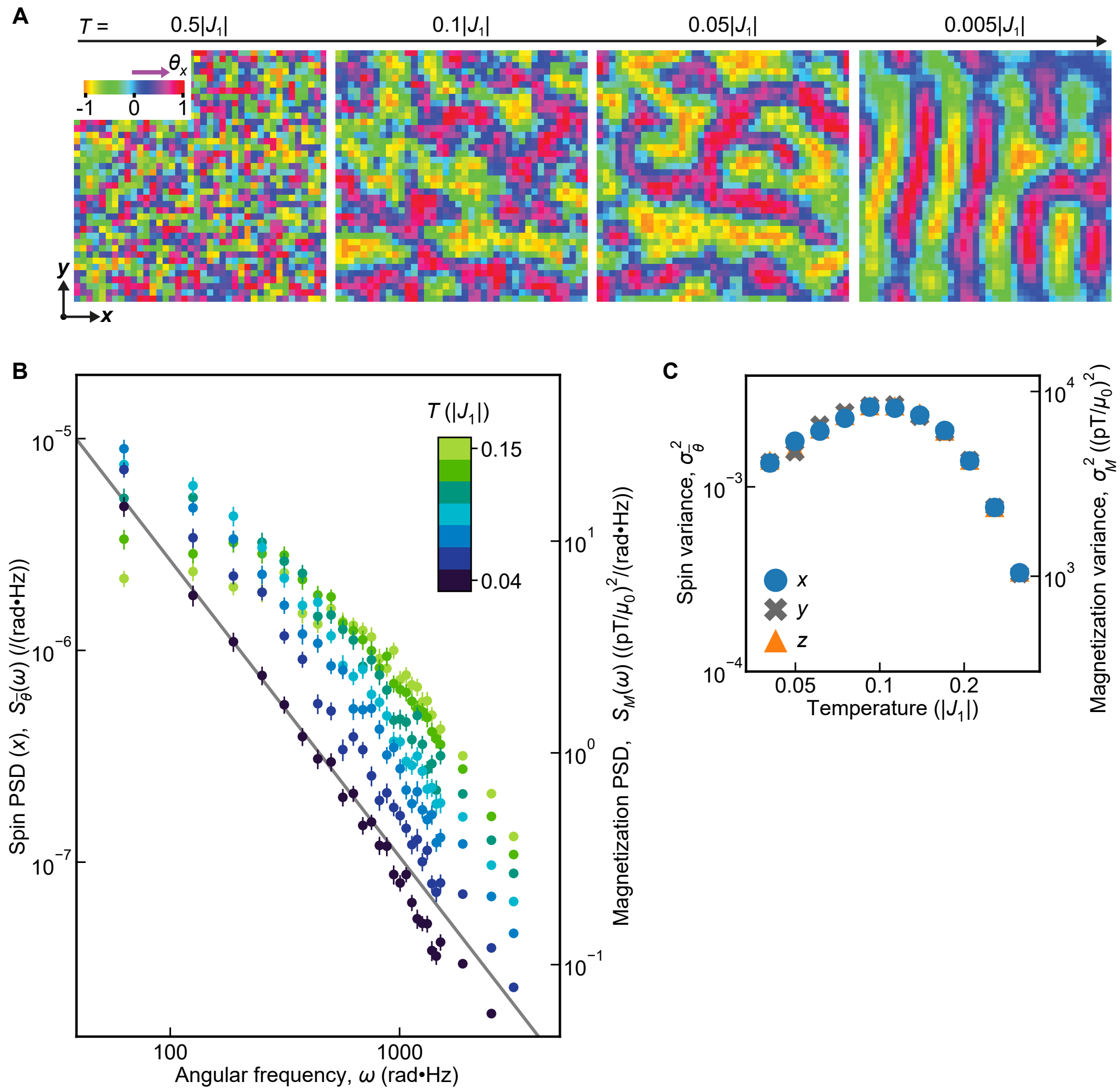

**Fig. S4. MC simulations of spiral spin liquid noise for Heisenberg spins.**

**A.** MC simulations of a snapshot of $\theta_x(\boldsymbol{r})$ on the SSL model of equation (1) using a square lattice with $N = 40 \times 40$ sites $\boldsymbol{r}$, each site with a three-dimensional spin unit vector $\boldsymbol{\vartheta}(\boldsymbol{r})$. Each snapshot is for a different temperature so that this sequence of SSL simulation snapshots is for approximately $T = 0.5|J_1|$, $0.1|J_1|$, $0.05|J_1|$ and $0.005|J_1|$.

**B.** From the time sequences of average $x$-component spin $\bar{\vartheta}_x(t,T)$ in the three-dimensional-spin simulation, the power spectral density of simulated noise $S_{\bar{\vartheta}_x}(\omega,T)$ is derived as a function of temperature $T$ and shown for seven selected temperatures. Again we take 1 MC time step to be $\tau = 1\ \mu$s. Here the error bars are the standard error of separated segments (Methods). The anticipated power spectral density of magnetization noise $S_M(\omega,T)$ is shown on the right-hand axis as estimated from calculations described in this document. The spectrum shows a powerful low-frequency noise down to $\omega/2\pi = 10$ Hz with a diminishing power below $T \sim 0.1|J_1|$. $\omega^{-1.4}$ line (gray) is drawn as a guide to the eye.

**C.** From the time sequences $\bar{\vartheta}_{x,y,z}(t,T)$ (the frequency component above 1 kHz is filtered out), the variance of simulated SSL noise $\sigma^2_{\bar{\vartheta}_{x,y,z}}(T)$ is presented. The variance peaks around $T = 0.1|J_1|$ and diminishes approximately as $T^1$.

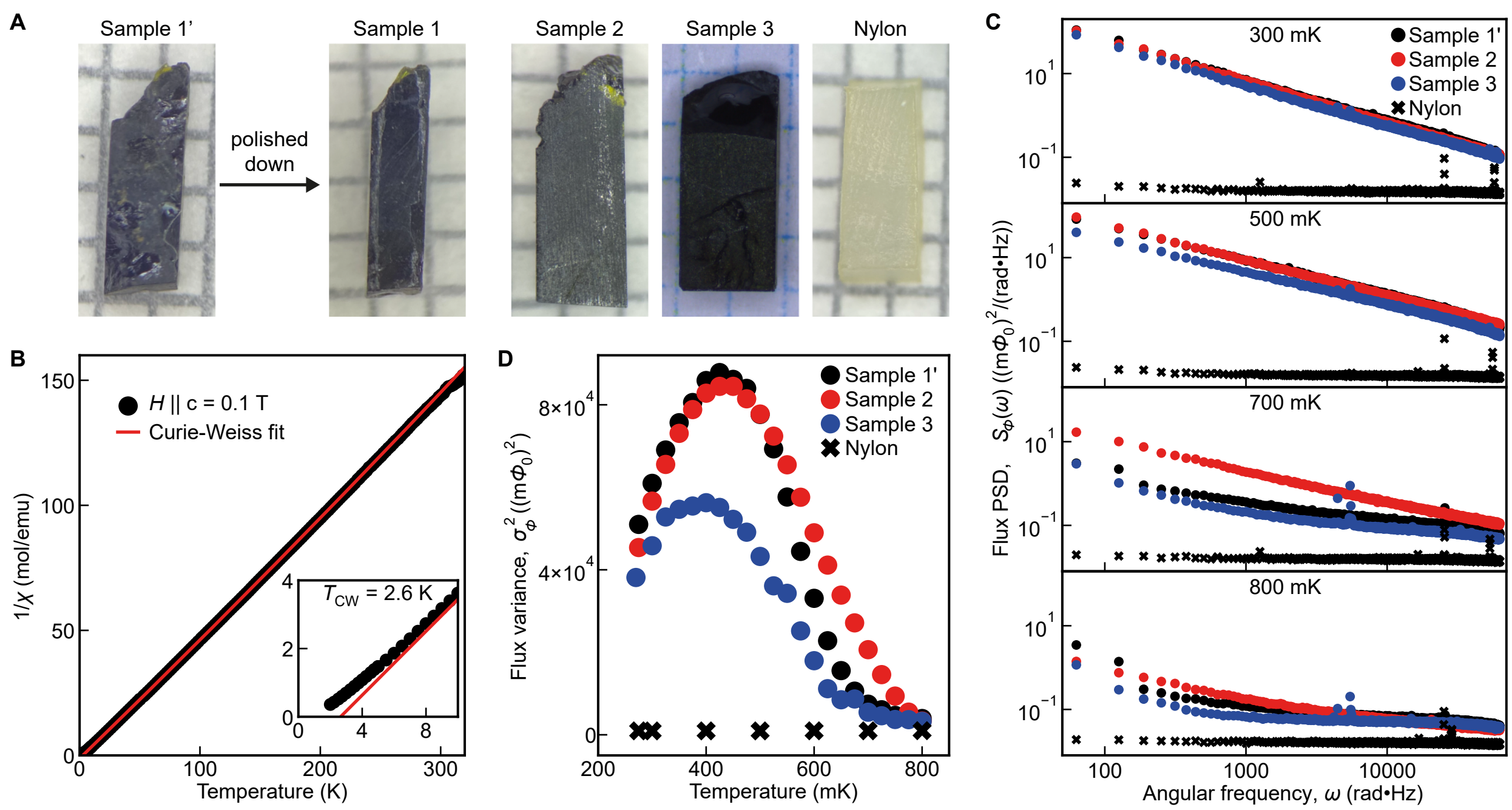

**Fig. S5. Single crystals of $Ca_{10}Cr_7O_{28}$ and comparison of their noise signals.**

**A.** Photos of the three $Ca_{10}Cr_7O_{28}$ samples and a nonmagnetic nylon block of a comparable size. Sample 1 is obtained by polishing down Sample 1'.

**B.** DC susceptibility of Sample 3 measured by MPMS. The Curie-Weiss fitting by $\chi = \chi_0 + \frac{C_{\mathrm{Curie}}}{T - T_{\mathrm{CW}}}$ (red line) yields $T_{\mathrm{CW}} = +2.6$ K and $C_{\mathrm{Curie}} = 2.1$ K•emu/mol corresponding to $\mu_{\mathrm{eff}} = 1.69\mu_{\mathrm{B}}$.

**C.** Comparison of the flux noise power spectral density $S_\Phi(\omega, T)$ of the three $Ca_{10}Cr_7O_{28}$ samples and the nylon block at four temperatures. All the $Ca_{10}Cr_7O_{28}$ samples generate a strong magnetic noise above the Nylon signal corresponding to a background noise.

**D.** Comparison of the flux noise variance $\sigma^2_\Phi(T)$ of the three $Ca_{10}Cr_7O_{28}$ samples and the nylon block. All the $Ca_{10}Cr_7O_{28}$ samples show a peak at $T \sim 400$ mK.

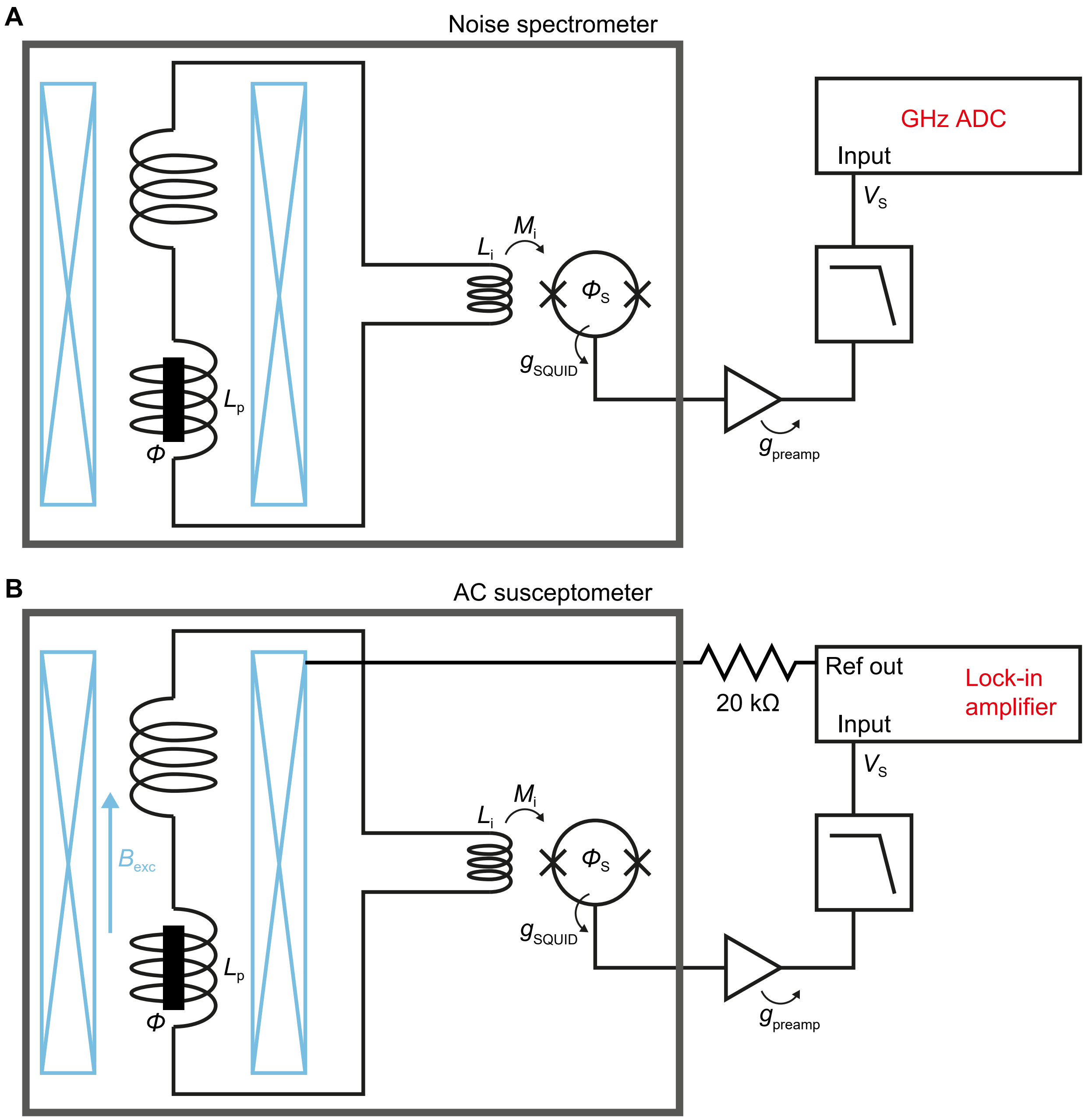

**Fig. S6. Schematic circuit diagrams of the experiment.**

**A.** Circuit diagram of the noise measurement.

**B.** Circuit diagram of the AC susceptibility measurement.

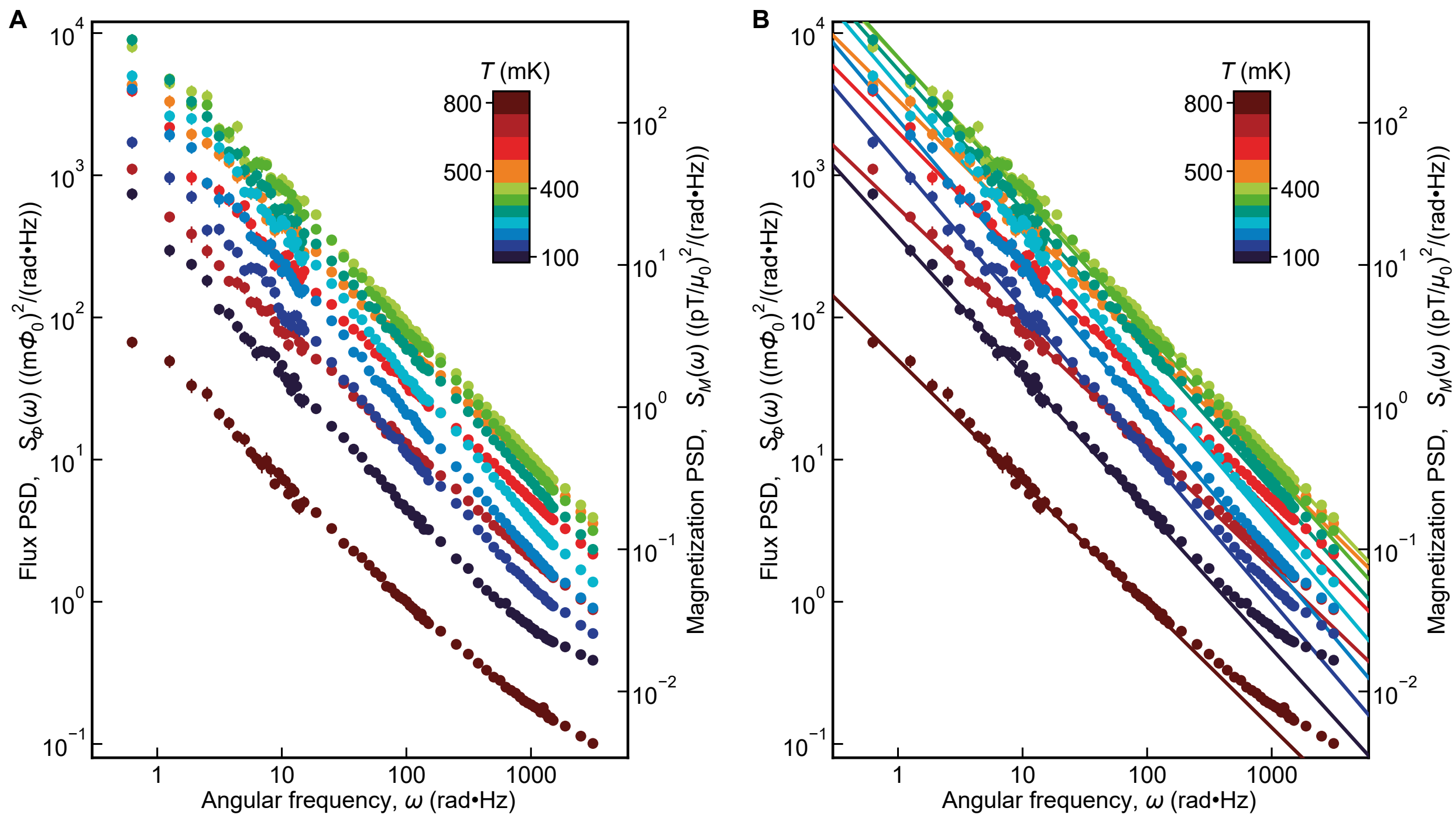

**Fig. S7. Experimental spin noise power spectral density of $Ca_{10}Cr_7O_{28}$ $S_\Phi(\omega, T)$ and fitting by $A(T)\omega^{-\alpha(T)}$.**

**A.** Experimental power spectral density of $Ca_{10}Cr_7O_{28}$ $S_\Phi(\omega, T)$ in the full measured-temperature range of $100 \text{ mK} \leq T \leq 800 \text{ mK}$. The equivalent power spectral density of magnetic field noise at the sample $S_M(\omega, T)$ is presented on right hand axis.

**B.** Fitting of experimental power spectral density $S_\Phi(\omega, T) = A(T)\omega^{-\alpha(T)}$ in the frequency range $0.1 \text{ Hz} \leq \omega/2\pi \leq 20 \text{ Hz}$ for $100 \text{ mK} \leq T \leq 800 \text{ mK}$.

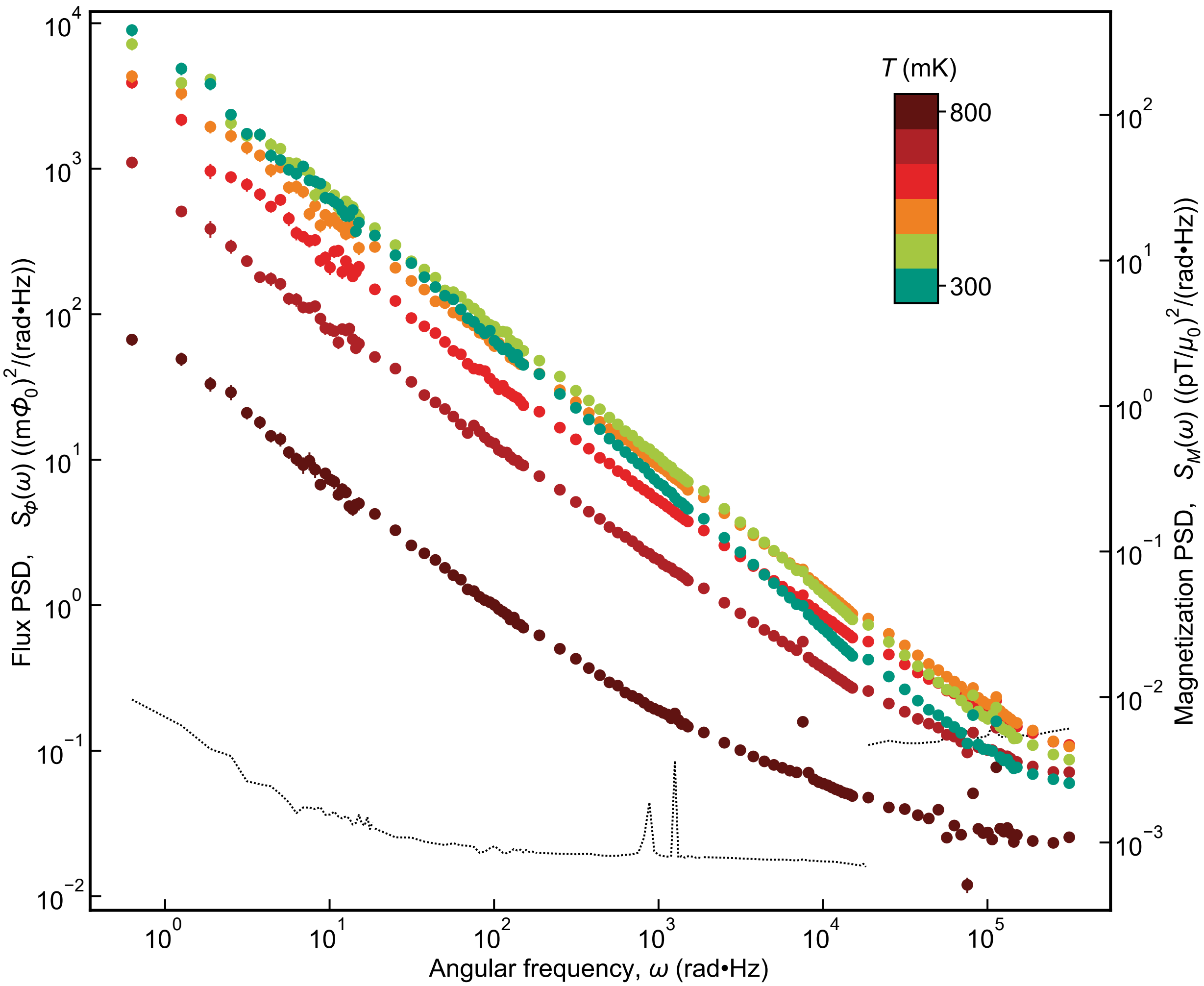

**Fig. S8. Experimental spin noise power spectral density of $Ca_{10}Cr_7O_{28}$ in an extended frequency range.**

Experimental power spectral density of $Ca_{10}Cr_7O_{28}$ in an extended frequency range of $0.1\ \text{Hz} \leq \omega/2\pi \leq\ 50\ \text{kHz}$ for six temperatures. The background noise measured for a nonmagnetic Nylon is plotted as a black-dotted line. The frequency range $\omega/2\pi$ above 3 kHz is from an extended-frequency measurement that enhances the noise floor. Above 3 kHz, $Ca_{10}Cr_7O_{28}$ noise is plotted after a subtraction by the background noise.

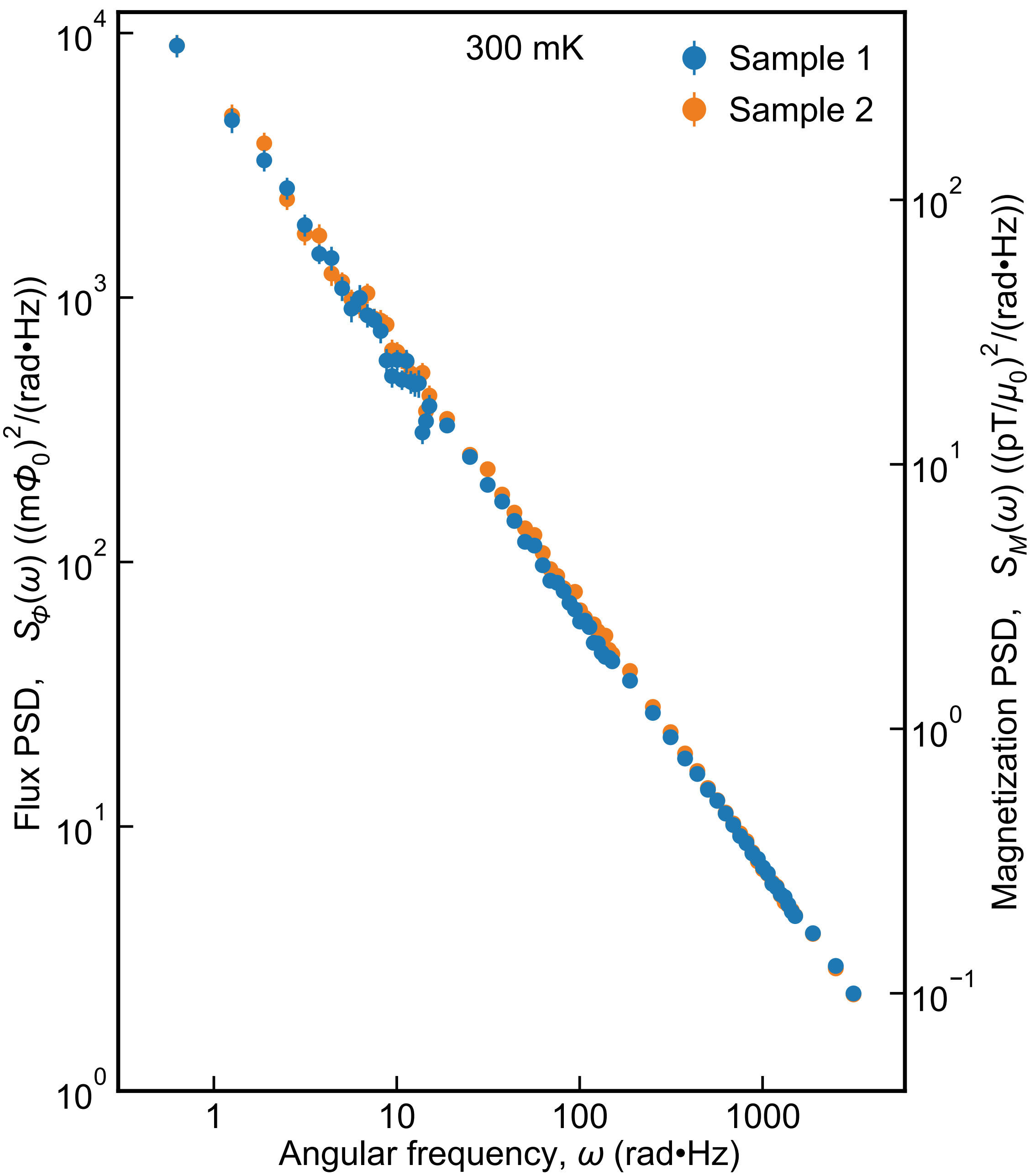

**Fig. S9. Comparison of experimental power spectral density from Sample 1 and Sample 2.**

The comparison of power spectral density $S_\Phi(\omega, T)$ from Sample 1 and Sample 2. $S_\Phi(\omega, T)$ of Sample 2 is scaled by $0.95^2$ so that the value of two flux power spectral densities agrees. The equivalent spontaneous magnetization noise $S_M(\omega, T)$ on the right axis is converted using the geometry of Sample 1.

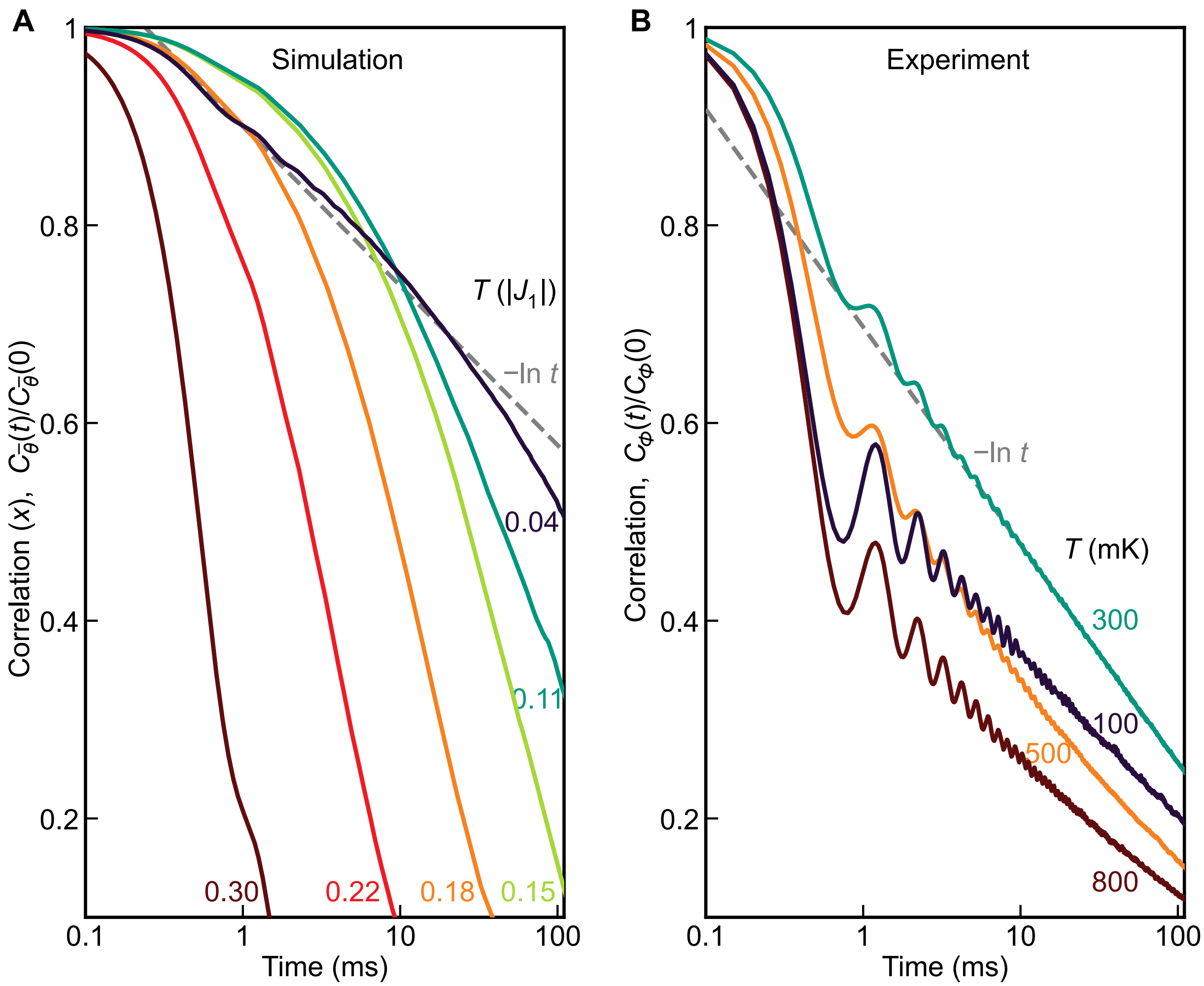

**Fig. S10. The correlation functions in ln *t* scale.**

**A.** The simulated correlation function $C_{\bar{\vartheta}_x}(t,T)/C_{\bar{\vartheta}_x}(0,T)$ of $x$-component spin noise (same quantity as Fig. 2D), plotted with a log scale horizontal axis. The correlation function gradually approaches to an almost linear behaviour $\propto -\ln t$ (gray dashed line) at low temperature.

**B.** The experimental correlation function $C_{\Phi}(t_k,T)/C_{\Phi}(0,T)$ of flux noise $\Phi(t)$ (same quantity as Fig. 4B), plotted with a log scale horizontal axis. The correlation function shows a linear behaviour$\propto -\ln t$ (gray dashed line).

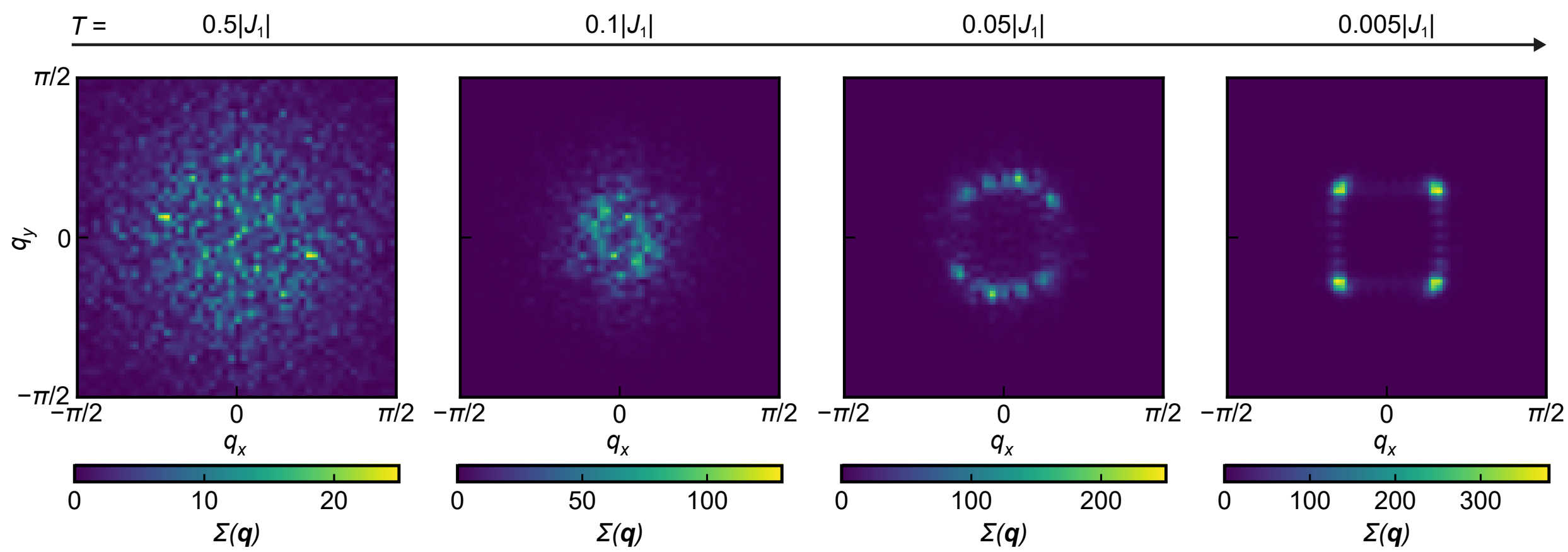

**Fig. S11. Structure factor of spiral spin liquid Monte Carlo simulations**

Structure factor $\Sigma(\boldsymbol{q})$ of the spiral spin liquid simulation, which is calculated from each spin pattern shown in Fig. 2A corresponding to $T = 0.5|J_1|$, $0.1|J_1|$, $0.05|J_1|$ and $0.005|J_1|$.

**Movie S1. Visualization of the spin equilibration process in an $L = 100$ system.**

The evolution of spin configuration as the $L = 100$ system is equilibrated by a total of $6 \times 10^6$ MC steps. The color of the pixel represents the direction of the spin with the same color code as Fig. 2A. $10^4$ MC steps are performed between each picture frame.

**Movie S2. Visualization of the spin equilibration process in an $L = 40$ system.**

The evolution of spin configuration as the $L = 40$ system is equilibrated by a total of $6 \times 10^6$ MC steps. The color of the pixel represents the direction of the spin with the same color code as Fig. 2A. $10^4$ MC steps are performed between each picture frame.